**From Novelty to Normalisation: Tracking Changing Perceptions of AI in Higher Education, 2024–2026**

Juliana Gerard*[a], Morgan Macleod[a], Kelly Norwood[b], Aisling Reid[c], Muskaan Singh[d]

[a] School of Communication and Media, Ulster University, Belfast, Northern Ireland

[b] School of Psychology, Ulster University, Coleraine, Northern Ireland

[c] School of Arts, English and Languages, Queen's University Belfast, Belfast, Northern Ireland

[d] School of Computing, Engineering and Intelligent Systems, Ulster University, Derry/Londonderry, Northern Ireland

**Abstract:** The rapid integration of generative artificial intelligence (AI) has reshaped the landscape of higher education. Students have embraced tools such as ChatGPT with striking speed, while teaching staff and institutions have responded with greater caution. Existing research on AI perceptions has mainly been cross-sectional, providing single-point snapshots that view attitudes as stable rather than evolving. This paper presents a longitudinal study of AI perceptions in higher education, tracking undergraduates, doctoral researchers, teaching staff and non-teaching staff at Ulster University across three survey waves between 2024 and 2026 (n=1,665). A quantitative survey design measured familiarity, reported use and perceived risk; results show that students rapidly normalised AI use over the period, moving from tentative experimentation to routine engagement, while staff expressed persistent concerns about academic integrity, assessment design, and critical thinking. Doctoral and non-teaching staff occupied intermediate positions, reflecting both pragmatic adoption and institutional caution. The student–staff gap widened as institutional policy struggled to keep pace with actual practice. By tracking these shifts directly rather than reconstructing them from disconnected studies, the paper moves beyond descriptive accounts of AI attitudes and demonstrates the importance of capturing perceptions in real time. The findings carry significant implications for adaptive institutional policy, AI literacy initiatives, and targeted staff training.

## 1. Introduction

Generative artificial intelligence (AI) has rapidly reshaped higher education. Tools such as ChatGPT have been adopted by students at a striking pace, while teaching staff and institutions have responded with a mixture of curiosity, caution, and regulatory concern. The visibility of AI in everyday academic practice continues to grow, raising urgent questions about its role in teaching, assessment, and knowledge production. Research into AI perceptions in universities has so far tended to treat attitudes as static; most studies provide a single snapshot of how students or staff feel at a given point in time.

Generalising from a single data point neglects how quickly technological change unfolds. Students' familiarity with AI tools is increasing year on year, while staff are grappling with how to address such tools in their teaching. Institutional guidance has shifted from prohibition and warnings toward regulation and responsible integration. Perceptions evolve as technologies are normalised, as policies adapt, and as the gap between student practice and staff expectations grows more visible.

A longitudinal perspective is therefore needed to understand how perceptions of AI in higher education develop over time. By tracking attitudes across multiple years, the present study reveals where consensus exists between students, doctoral researchers, teaching staff, and support staff, and where differences between these groups are widening or narrowing. The findings provide evidence for how universities might respond through training, policy development, and the design of learning environments that reflect the realities of AI's presence in academic life.

## 2. Literature review

### 2.1 AI policy in education contexts

#### *2.1.1 AI policy in Higher Education*

The release of ChatGPT in late 2022 prompted rapid and uneven responses across UK higher education. Early institutional reactions in 2023 ranged from outright prohibition to cautious openness. The Department for Education (2023) published an evidence summary acknowledging concerns about academic misconduct, over-reliance and data security. The Quality Assurance Agency followed up with advice on maintaining standards in May 2023 and extended its guidance in July with a report that reconsidered assessment in the ChatGPT era (QAA, 2023a, 2023b). The Russell Group (2023) published principles prioritising integrity, transparency, and staff–student collaboration. JISC developed a generative AI primer alongside student perception reports tracking adoption patterns from 2023 onwards (JISC, 2023, 2025).

By 2024, the sector had shifted from prohibition toward responsible integration. The QAA's Quality Compass (2024) signalled that blanket bans were neither effective nor enforceable, encouraging assessment redesign and AI literacy development. Wilson's (2025) analysis of Russell Group policies confirms that member institutions moved from initial caution to elaborated frameworks, though implementation remained inconsistent. By 2025, institutional energy had shifted from detection toward guidance, disclosure, and curriculum redesign.

#### *2.1.2 AI policy at Ulster University*

Ulster University's response followed the broader UK trajectory while adopting a notably integrative stance from an early stage. Student guidance published in August 2023 declined to prohibit AI tools outright, positioning Ulster among institutions that choose to explore their use in 'ethical, transparent and reasonable ways' (Ulster University, 2023). The Academic Misconduct Policy was updated in 2023–24 to explicitly name AI-generated content and require students to acknowledge any such use in submission declarations.

UU’s approach is underpinned by the institutional strategy 'People, Place and Partnership: Delivering Sustainable Futures for All', which frames AI through the values of inclusion, wellbeing and active thinking (Ulster University, 2024). Separate guidance for postgraduate researchers and supervisors was developed in parallel, acknowledging that AI raises distinct questions at doctoral level. Ulster hosted the National Conference on Generative Artificial Intelligence in Education in 2024, positioning the university as an active participant in sector-level debate. The policy environment between 2023 and 2025 thus evolved from initial acknowledgement toward a more developed framework for responsible use, providing the institutional context within which the perceptions tracked in the present study developed.

### 2.2 Research on AI perceptions

A substantial body of research has accumulated on how students and staff in higher education perceive generative AI. Methods range from large quantitative surveys (Chan & Hu, 2023) to small-scale qualitative studies with specific cohorts (Alshamy et al., 2025). Several patterns recur. Students perceive AI tools as beneficial for writing support and academic performance. Staff perspectives are far more cautious. Teaching staff voice persistent concerns about academic integrity and assessment design, as well as fears that AI displaces critical thinking.

The divergence between student enthusiasm and staff caution is well documented. Chan and Hu (2023) found that students expressed broadly positive attitudes toward AI tools while flagging concerns about accuracy and over-reliance. Alshamy et al. (2025) drew on responses from 555 students and 168 academics at Sultan Qaboos University. Notable differences emerged across aspects of AI acceptance: students viewed usefulness more positively, while staff worried about integrity. JISC's reports on student perceptions (2023, 2025) reinforce these findings in a UK context, recording appetite for institutional guidance on ethical use rather than prohibition.

#### *2.2.1 Changes in AI perceptions over time*

Read in sequence, the studies suggest significant shifts. Early investigations documented novelty and uncertainty, as students experimented with AI tools in the absence of clear institutional guidance (Chan & Hu, 2023). Subsequent research describes greater familiarity and more routinised use, with the student–staff divergence becoming more visible over time (Alshamy et al., 2025). Sector-wide surveys reinforce the trajectory: Tyton Partners (2023) found that nearly half of undergraduates used AI regularly, compared with one-fifth of faculty. JISC's reports (2023, 2025) come closest to a longitudinal perspective in a UK context, tracking patterns across successive years.

#### *2.2.2 The longitudinal gap*

A persistent limitation of the existing research is its cross-sectional character. Most studies collect data at a single point in time, treating perceptions as stable rather than evolving. Even mixed-methods research typically acknowledges temporal change only in its limitations. The overall arc from novelty to normalisation must therefore be inferred indirectly through disconnected studies. Doctoral researchers and professional services staff remain largely absent from the published literature, which concentrates on undergraduates or teaching faculty. The present study addresses these gaps directly by tracking attitudes across three survey waves at Ulster University between 2024 and 2026.

## 3. Methods

### 3.1 Participants

Participants were recruited from across Ulster University, encompassing undergraduate and postgraduate taught students, doctoral researchers, teaching staff and non-teaching staff (total N = 1,665), with participation numbers varying marginally by year (Table 1). This broader inclusion was intended to address a gap in existing research, which tends to focus primarily on undergraduates and teaching staff. Data was collected in semester 2 (February) of three academic years (2023-2024, 2024-2025, 2025-2026).

A new sample was recruited for each year, although participants who had completed the study previously were permitted to repeat the survey in subsequent years. Undergraduate students formed the largest group, followed by staff with teaching responsibilities, while doctoral and non-teaching staff provided smaller but important perspectives often absent from AI perception studies. The study was approved by the Ulster University Communication and Media Risk and Ethics Filter Committee, and participants received a £10 Amazon voucher for their participation.

Table 1. Participant numbers, by population and year (total N = 1,665)

| | **Year** | | | |
|---|---|---|---|---|
| **Population** | *2024* | *2025* | *2026* | ***Population total*** |
| *student* | 512 | 401 | 401 | **1,314** |
| *PhD* | 56 | 61 | 57 | **174** |
| *teaching staff* | 68 | 35 | 74 | **177** |
| *non-teaching staff* | 144 | 102 | 112 | **358** |
| ***Year total*** | **780** | **599** | **644** | **1,665** |

### 3.2 Design

The study employed a longitudinal survey design, which was completed remotely (online), combining repeated cross-sectional elements with a stable core of questions across years (Petricini 2024, Gerard et al 2025). Specifically, each survey consisted of statements about AI along with 5 Likert ratings (from 'Strongly Disagree' to 'Strongly Agree'), and participants were prompted to 'Select your level of agreement or disagreement' with each statement.

The full set of statements is presented in Table 2, categorised by core elements of AI perceptions (adapted from Petricini et al 2024):

- Familiarity/prior experience
- Use in practice

- Training
- AI and coursework (students and instructors)
- Institutional policy
- Student concerns (value/benefit and instructor use)

The statements were adapted from Petricini et al. ’s (2024) study on student and staff perceptions across subject areas and addressed participants’ familiarity with generative AI, prior experience, perceived benefits and risks, and how AI was used in practice (e.g., for coursework, teaching materials, or marking). Additional items captured the degree to which institutional policy was visible or influential at each time point. Wording varied slightly across populations to reflect respective experience - e.g. AI in coursework/modules for students vs in teaching for staff. To ensure comparability, core measures were retained across all three years, with only minor changes to the survey framing in the second and third waves to reflect the repeated delivery.

Table 2. Statements on the survey, by category - rated on a 5-point scale from ‘Strongly Disagree’ to ‘Strongly Agree’

| **Category** | **Statement** |
|---|---|
| Familiarity/ prior experience | I am familiar with the concept of artificial intelligence (AI) |
| | I am familiar with ChatGPT or other AI text generation tools |
| | I have experience using ChatGPT or other text generation tools |
| Use in **practice: students** | My instructors have addressed the use of AI in my modules (student) |
| | I have addressed the use of AI with my supervisor/students/team |
| | My instructors have integrated AI generators like ChatGPT into their instruction (student) |
| | I integrate AI tools in my research/instruction/job |
| | I plan to use ChatGPT or similar tools for my coursework in the future (student) |
| | I plan to integrate ChatGPT and/or other AI text or image generators into my research/instruction/job in the future |
| Training | I have received instructions about how to use ChatGPT or similar tools (student)<br>I have participated in training on the use of ChatGPT and/or similar AI text and image generators<br><br>I would be open to receiving instructions about how to use ChatGPT or similar tools (student);<br>I am interested in receiving training on the use of ChatGPT and/or similar AI text and image generators |
| AI and coursework: students | Students' use of AI text generation tools to complete coursework is prevalent in higher education<br><br>Students' use of AI text generation tools to complete coursework is inevitable<br><br>Students should be restricted from using AI for coursework |
| AI and coursework: instructors | Instructors' use of AI text generation tools to complete coursework is prevalent in higher education<br><br>Instructors' use of AI text generation tools to complete coursework is inevitable |
| Institutional policy | The use of AI text-generation tools could be beneficial for students<br><br>I trust AI in marking my assignments and assessments for my modules instead of my instructor (student)<br>I trust AI in marking students' assignments and assessments<br><br>Use of AI text generation tools to complete coursework is inconsistent with academic integrity policies at the University<br><br>Students will need to be taught how to use AI text generation tools appropriately |
| Student concerns<br>(students only): Value/benefit | Artificial Intelligence has value in education |
| | AI is used in education for good and helpful reasons |
| | Instructors are confident in their use of AI in academic settings |
| Student concerns (students only): instructor use | I would feel confident knowing an instructor was using an AI-created teaching resource |
| | I would want to be informed if my instructor was using AI-created resources on courses (students only) |

### 3.3 Analysis plan and predictions

Quantitative data were analysed descriptively and inferentially to track changes over time and across groups. Analyses compared mean ratings to each question by converting the Likert scale responses to numeric ratings from 1 (‘Strongly Disagree’) to 5 (‘Strongly Agree’). If perceptions have changed over time, then the mean ratings for the statements in Table 2 are expected to differ across populations and time points (2024, 2025, and 2026). However, changes in perceptions may also be reflected in just some of the perception measures (not all of them), or for just some of the populations, or between two time points rather than all three (e.g. from 2024 to 2025, but not in the final year).

To explore changes in ratings over time, the first step in the analysis is a linear regression model including the variables ‘population’, ‘year’, and ‘question’, along with their interactions, with Likert ratings as the dependent variable. Where different changes are observed over time across populations, significant interactions are expected between these factors. However, to probe the precise nature of these changes, pairwise posthoc tests are used to compare individual means, providing both trend data and nuanced insight into how perceptions of AI evolved between 2024 and 2026.

## 4. Results

Significant effects for all questions are presented in Table 3, by population and year; the effects were identified via the steps described in Section 3.3, specifically:

1. A mixed effects linear regression using R (R core team) with lmer (Bates et al 2015) with rating as the dependent measure, fixed effects ‘population’, ‘year’ and ‘question’ and ‘subject’ as a random effect (intercepts only).
2. Posthoc pairwise comparisons using R with emmeans: for each question in each population, compare mean ratings between timepoints.

The posthoc tests first provide a picture of overall changes for the duration of the study, with a comparison of the first and last time points (i.e. 2024 and 2026). Next, comparisons were conducted between time points, i.e. between 2024 and 2025, and between 2025 and 2026.

Table 3. β values for significant differences (+ for positive change, - for negative change)

| | | student | | | PhD | | | teaching | | | non-teaching | | |
|---|---|---|---|---|---|---|---|---|---|---|---|---|---|
| | | 2024 - 2026 | 2024 - 2025 | 2025 - 2026 | 2024 - 2026 | 2024 - 2025 | 2025 - 2026 | 2024 - 2026 | 2024 - 2025 | 2025 - 2026 | 2024 - 2026 | 2024 - 2025 | 2025 - 2026 |
| Familiarity/prior experience | I am familiar with the concept of artificial intelligence (AI) | 0.25 | | | | | | | | | | | |
| | I am familiar with ChatGPT or other AI text generation tools | 0.28 | | 0.19 | | | | 0.49 | | | 0.59 | | 0.47 |
| | I have experience using ChatGPT or other text generation tools | 0.39 | 0.31 | | 0.58 | | | 0.75 | | | 0.87 | 0.36 | 0.51 |
| Use in practice | My instructors have addressed the use of AI in my modules (student);<br>I have addressed the use of AI with my supervisor/students/team | 0.78 | 0.42 | 0.36 | 1.11 | | 0.91 | 0.82 | | 0.80 | 1.14 | | 1.01 |
| | My instructors have integrated AI generators like ChatGPT into their instruction (student);<br>I integrate AI tools in my research/instruction/job | 0.59 | 0.36 | 0.23 | 0.49 | | | 0.82 | | | 0.86 | | 0.63 |
| | I plan to use ChatGPT or similar tools for my coursework in the future (student);<br>I plan to integrate ChatGPT and/or other AI text or image generators into my research/instruction/job in the future | 0.24 | 0.28 | | | | | | | | | | |
| Training | I have received instructions about how to use ChatGPT or similar tools (student);<br>I have participated in training on the use of ChatGPT and/or similar AI text and image generators | 0.76 | 0.33 | 0.43 | 1.39 | | 1.07 | 0.88 | | 1.14 | 0.44 | | 0.50 |
| | I would be open to receiving instructions about how to use ChatGPT or similar tools (student);<br>I am interested in receiving training on the use of ChatGPT and/or similar AI text and image generators | -0.39 | | -0.31 | | | | | -0.64 | | | | |

| | | student | | | PhD | | | teaching | | | non-teaching | | |
|---|---|---|---|---|---|---|---|---|---|---|---|---|---|
| | | 2024 - 2026 | 2024 - 2025 | 2025 - 2026 | 2024 - 2026 | 2024 - 2025 | 2025 - 2026 | 2024 - 2026 | 2024 - 2025 | 2025 - 2026 | 2024 - 2026 | 2024 - 2025 | 2025 - 2026 |
| AI and coursework: Students | Students' use of AI text generation tools to complete coursework is prevalent in higher education | 0.48 | 0.30 | | | | | 0.62 | | | 0.47 | | |
| | Students' use of AI text generation tools to complete coursework is inevitable | | | | | | | | | | | | |
| | Students should be restricted from using AI for coursework | 0.25 | | 0.26 | -0.50 | -0.56 | | | | | | | |
| AI and coursework: Instructors | Instructors' use of AI text generation tools to complete coursework is prevalent in higher education | | | | | | | 0.47 | | | 0.34 | | 0.38 |
| | Instructors' use of AI text generation tools to complete coursework is inevitable | | | | | | | | | | | | |
| Institutional policy | The use of AI text-generation tools could be beneficial for students | | | | | | | | | | | | |
| | I trust AI in marking my assignments and assessments for my modules instead of my instructor (student);<br>I trust AI in marking students' assignments and assessments | | | | | | | | | | | | |
| | Use of AI text generation tools to complete coursework is inconsistent with academic integrity policies at the University | | | | | | | | | | | | |
| Student concerns: Value | Artificial Intelligence has value in education | - 0.26 | | -0.28 | | | | | | | | | |
| | AI is used in education for good and helpful reasons | - 0.20 | | -0.26 | | | | | | | | | |
| Student concerns: Instructor use | Instructors are confident in their use of AI in academic settings | | | | | | | | | | | | |
| | I would feel confident knowing an instructor was using an AI-created teaching resource | - 0.18 | | | | | | | | | | | |
| | I would want to be informed if my instructor was using AI-created resources on courses | | | | | | | | | | | | |

The results in Table 3 include the changes across the study duration, but the ratings themselves are needed to provide context for these changes. In the following sections, we explore these changes based on the categories defined in section 3.2: Familiarity/prior experience, Use in practice, Training, AI and coursework, Institutional policy and Student concerns (students only).

### 4.1 Familiarity/Prior experience

Results for familiarity/prior experience are presented in Figures 1-2. Mean ratings are high across all three statements, generally ranging between ‘Agree’ and ‘Strongly Agree.’ However, numerical increases are evident even within this range.

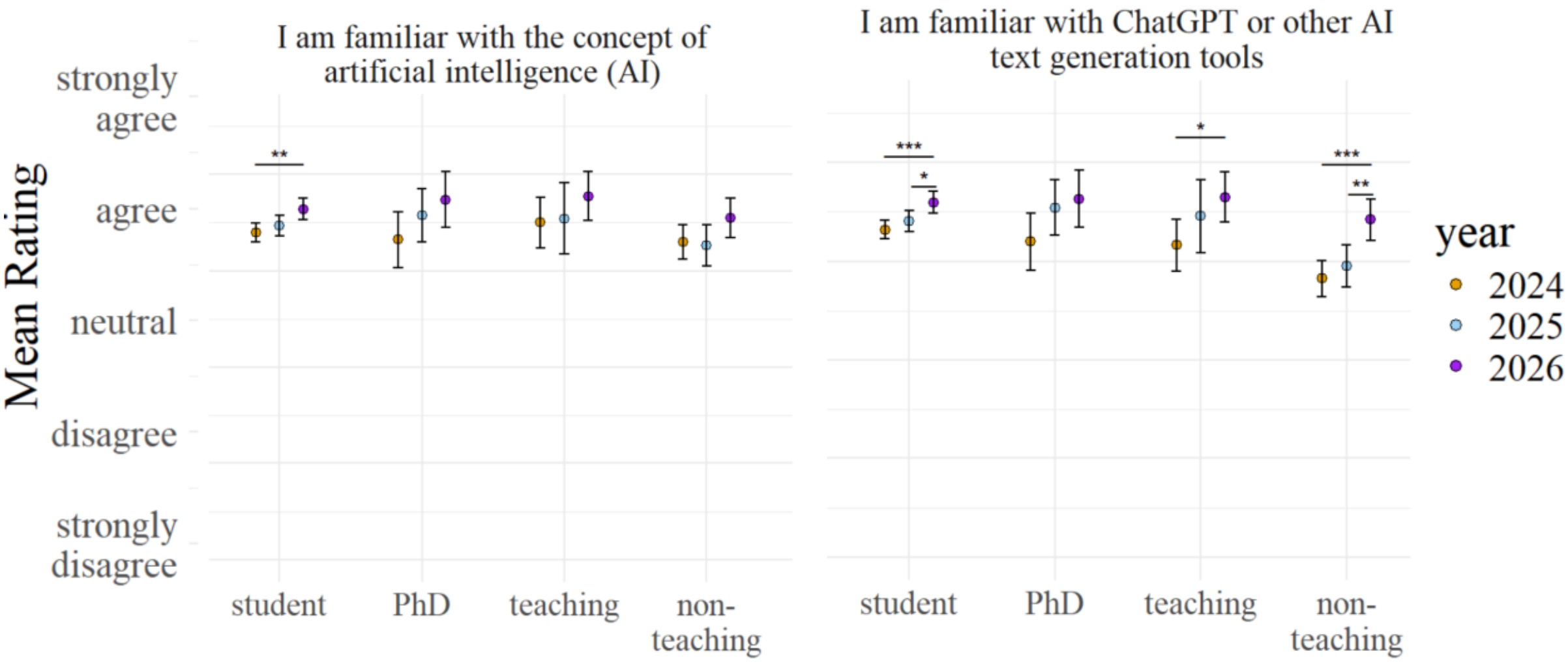

Figure 1 Figure 2

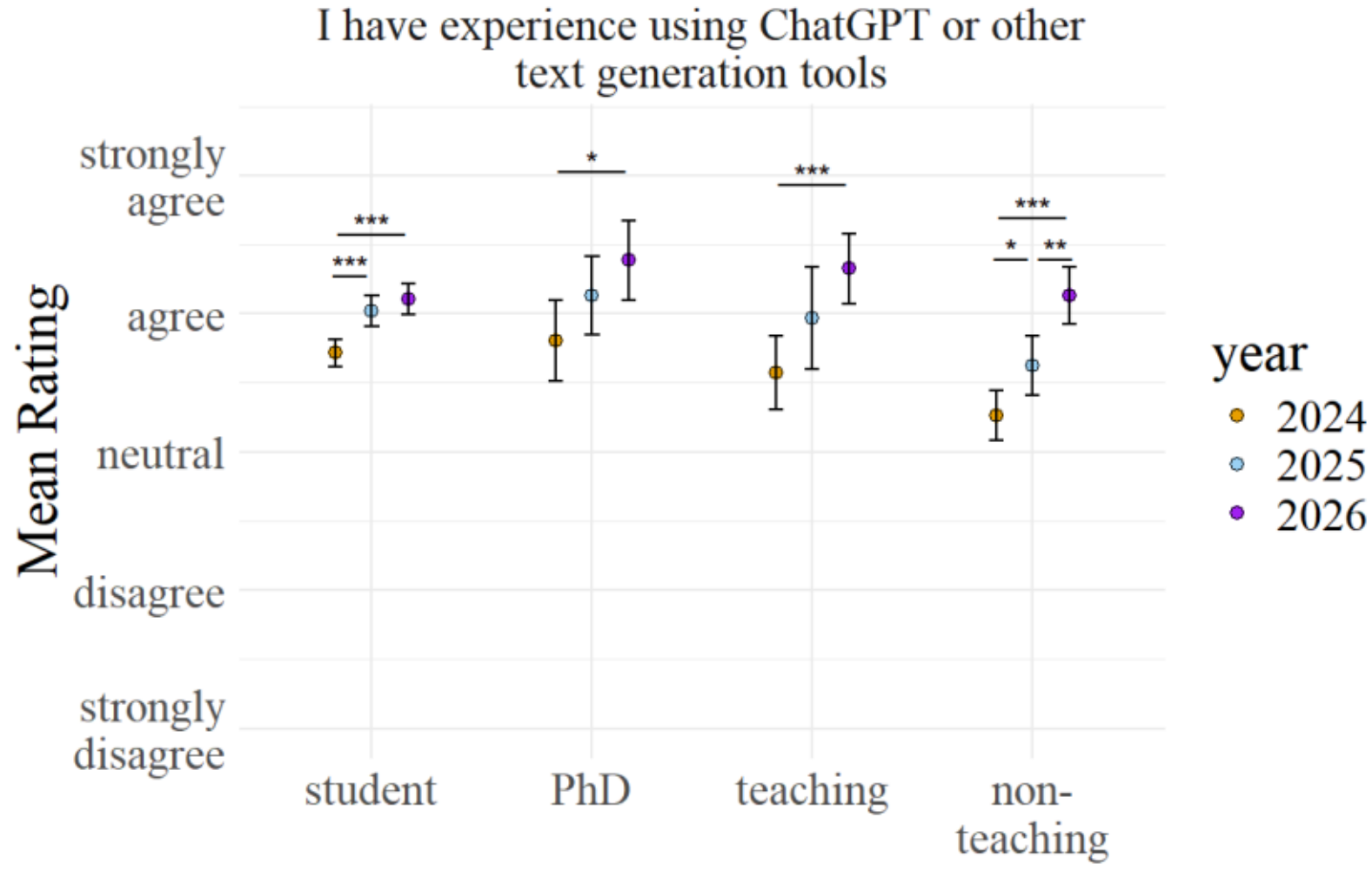

Figure 3

Notably, significant increases in familiarity (Figure 1) and experience (Figure 3) were observed for all populations, save for PhDs' familiarity (likely due to a smaller sample size). Moreover, the magnitude of these increases indicates a changing distribution over time. In fact, these shifts reflect shifts in a bimodal distribution: participants tended not to give a 'Neutral' response, but rather, increased experience resulted from decreases in 'Disagree' and 'Strongly Disagree' responses (Appendix 1).

**4.2 Use in practice**

Results for use in practice are presented in Figures 4-6, with a striking contrast between the patterns observed for actual changes in integration of AI and future plans to use AI. First, significant increases in addressing AI use were observed at both time points among students, reflecting an early effort to address AI use in modules. In contrast, this increase was observed only at the second time point for the other three populations.

Similar patterns are observed for the integration of AI; however, with far lower ratings at all time points: while the use of AI is addressed (Figure 4), it is less likely to be integrated into daily practice (Figure 5). These rates may also shed light on the lack of change observed in Figure 6: the ratings for integration in 2026 are similar to the ratings for planned use throughout all time points. We return to this observation and its implications in the discussion.

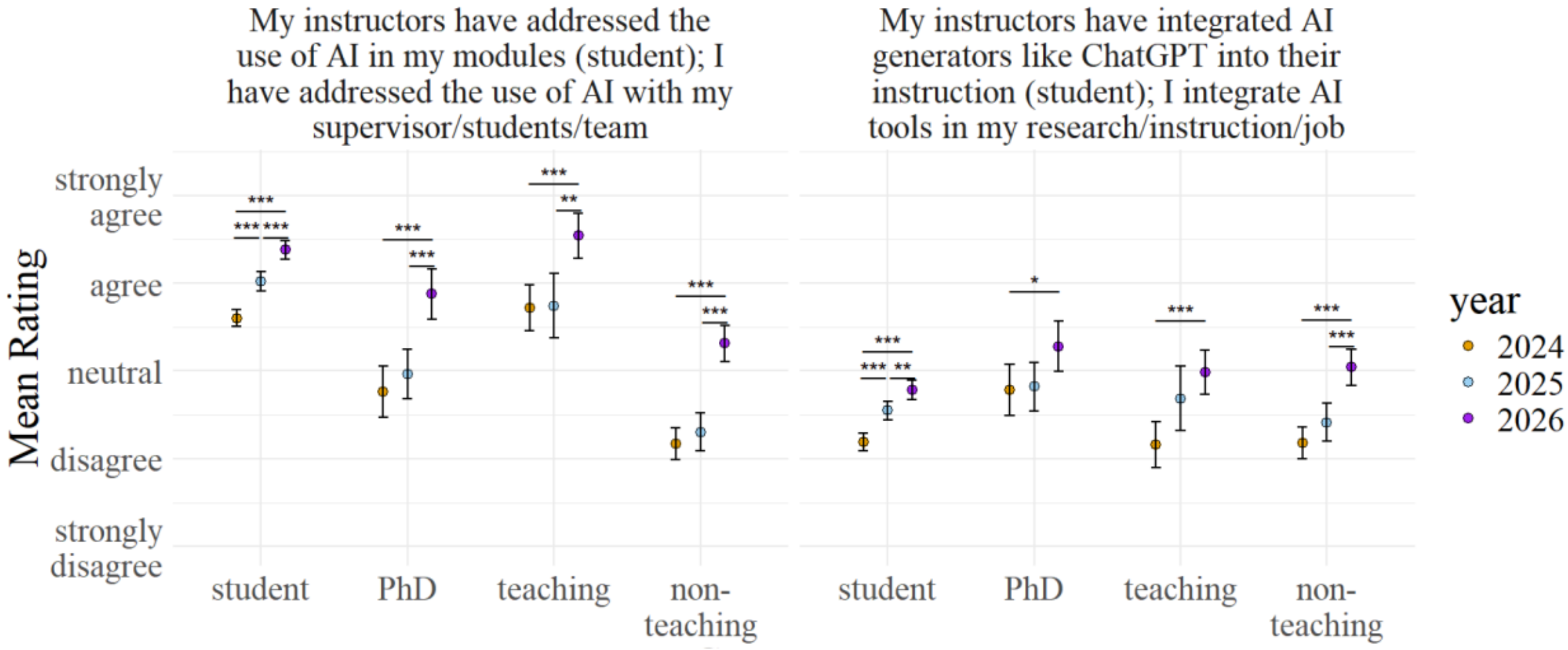

Figure 4

Figure 5

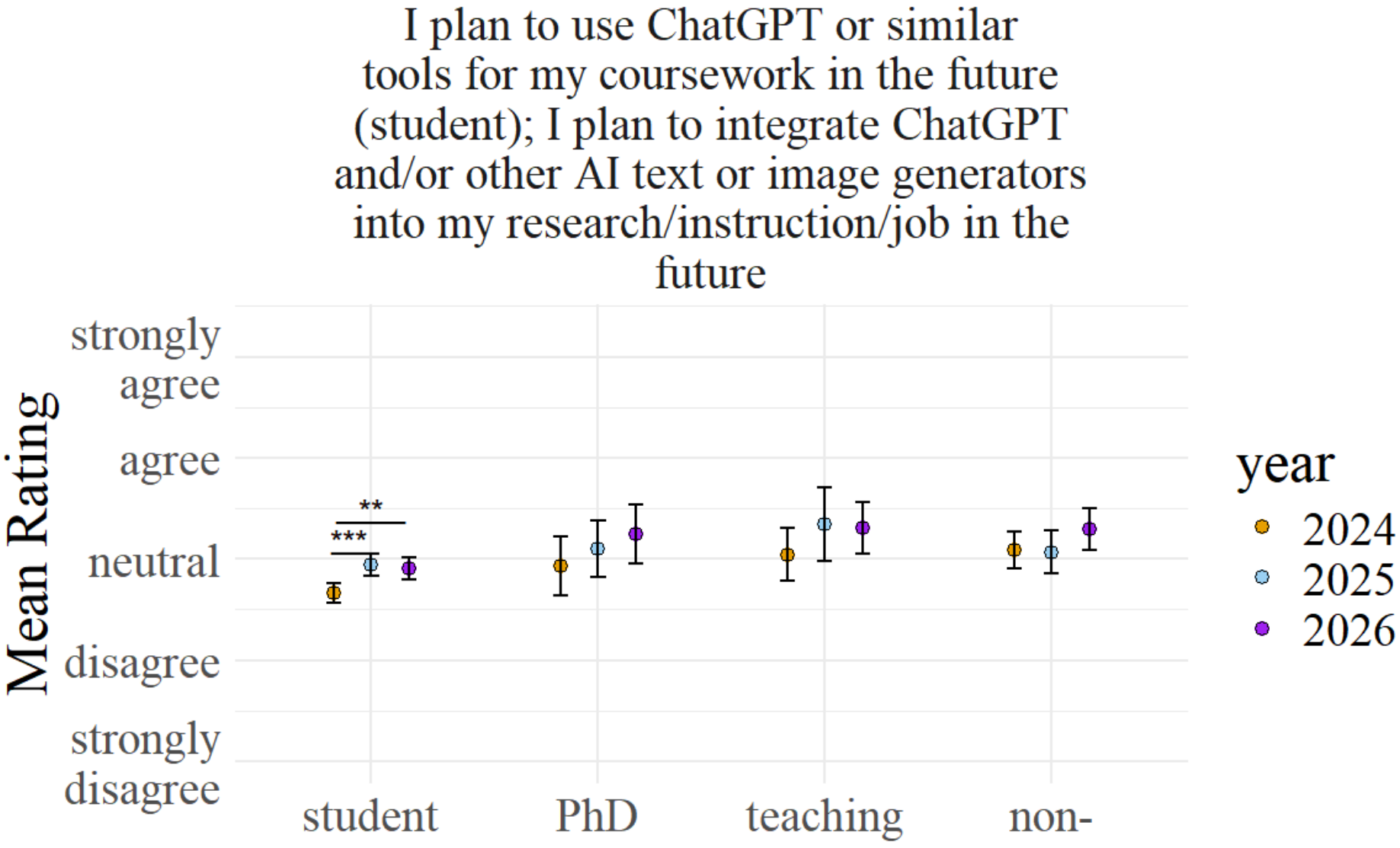

Figure 6

### 4.3 Training

Results for training are presented in Figures 7-8, and reflect an interesting parallel to the previous contrast between integration and plans to use AI: as for integration, training on AI has increased across all four populations (Figure 7). Notably, the same trend is observed as in several previous questions: this increase is seen across all timepoints for students, but only at the second timepoint (from 2025 to 2026) for the three other populations. Meanwhile, interest in training remained relatively stable and generally high, with the exception of students, whose responses decreased significantly at each timepoint (this decrease was also observed at the first timepoint for instructors). This result for students was unexpected, and we consider possible sources in the discussion.

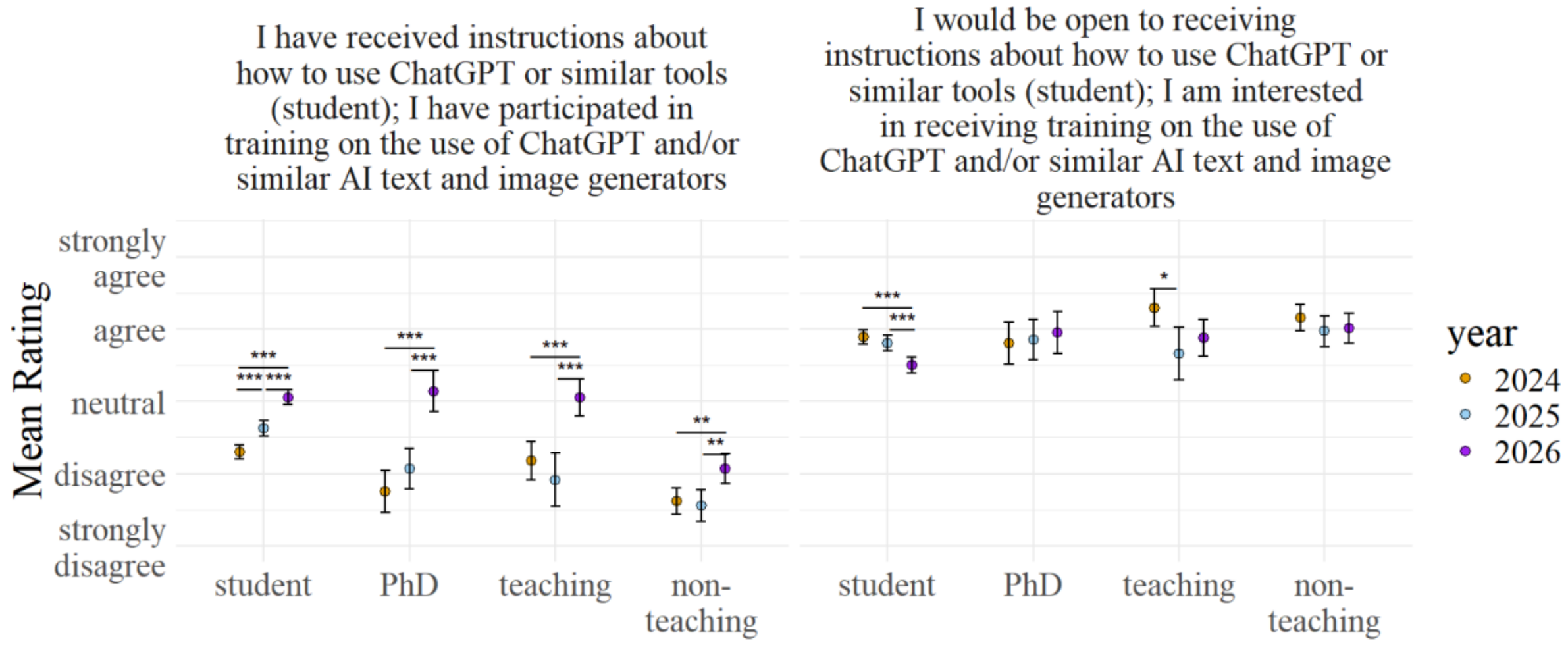

Figure 7

Figure 8

### 4.4 AI and coursework

The statements about AI and coursework included two sub-themes, addressing coursework and students, followed by coursework and instructors. We explore the results of each theme separately.

*4.4.1 Students*

Results for AI and coursework for students are presented in Figures 9-11, and reveal a mixed picture. Students increasingly agree that students' use of AI to complete coursework is prevalent, with a significant increase at the first timepoint. This same increase is observed numerically for the other three populations, but it is not significant, possibly an artefact of the high ratings for this item. In contrast, ratings for the statement that students' AI use is "inevitable" stayed constant for all four populations.

Results are mixed for the statement about restricting students' use of AI for coursework: students' ratings increased, while PhD researchers' ratings decreased, with no changes in ratings among teaching and non-teaching staff. Notably, all ratings are near a "neutral" mean across all timepoints, and for all populations at all timepoints, there was high variation in responses, with ratings distributed across all response values (see Appendix 1).

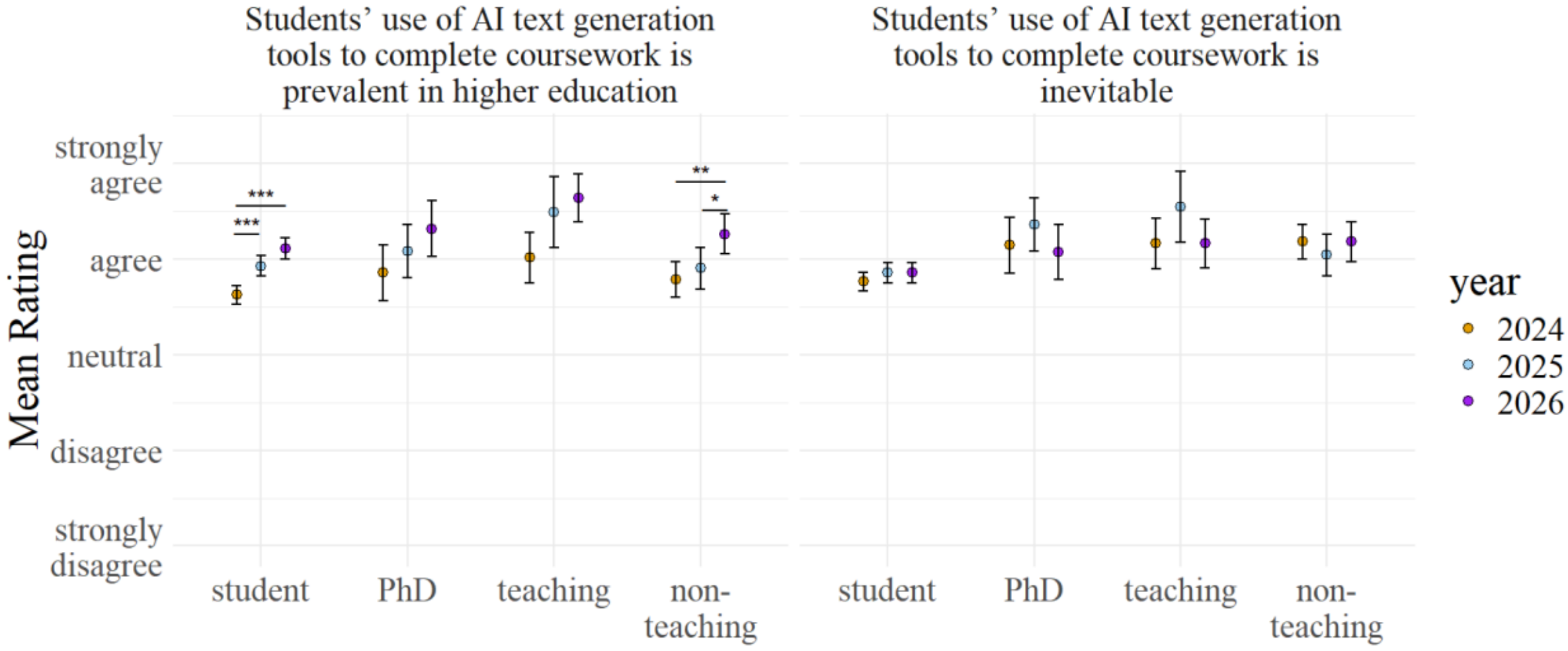

Figure 9

Figure 10

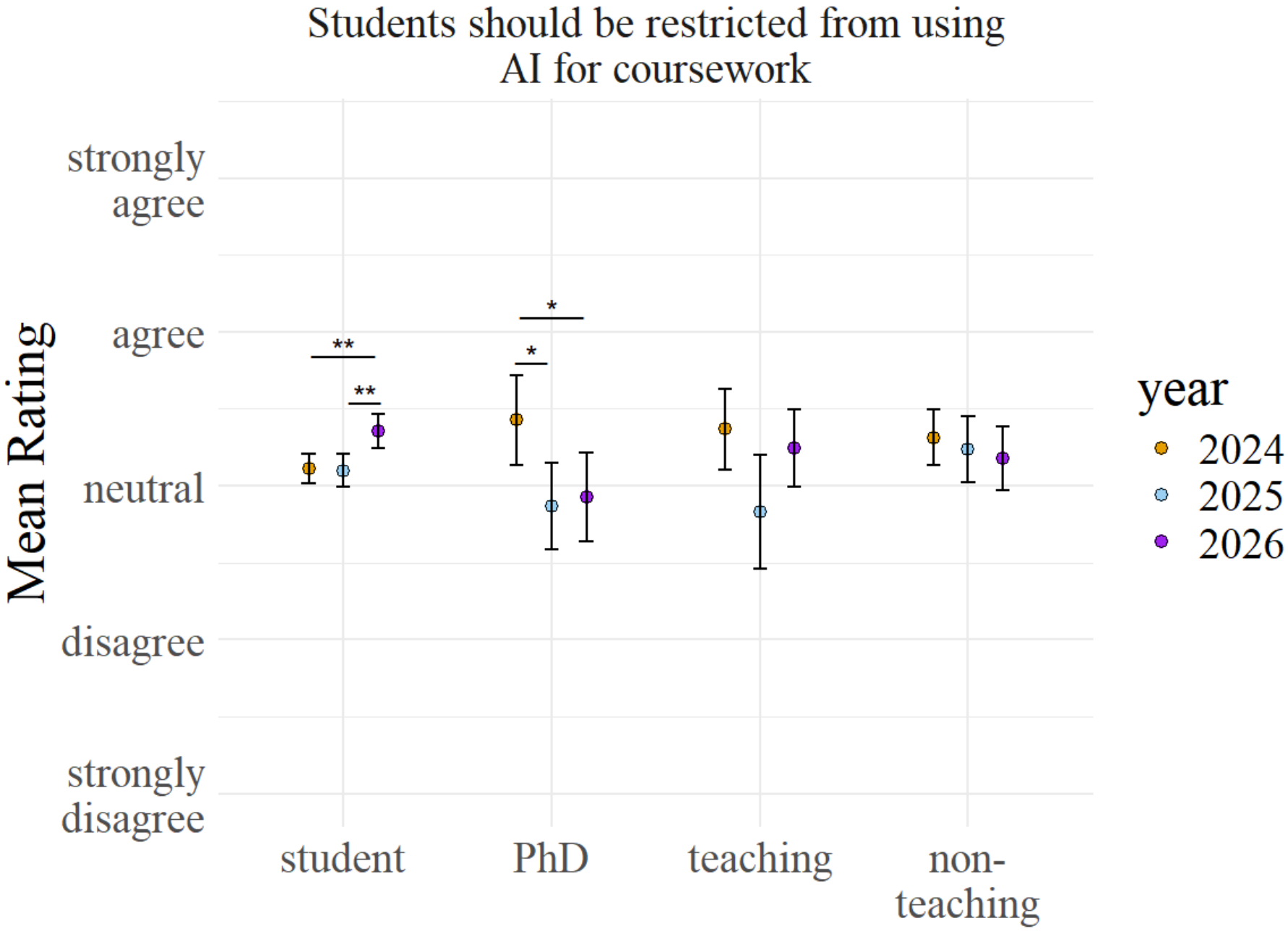

Figure 11

*4.4.2 Instructors*

Results for AI and coursework for instructors are presented in Figures 12-13. As for the student results, they present a mixed picture: teaching and non-teaching staff rate an increase in instructors' use of AI for coursework, with no change for PhD researchers. No change in ratings is observed for the inevitability of instructors' use of AI for coursework.

One notable contrast between the ratings for instructors' use and students' use is that the ratings for instructors' use are far lower. However, these ratings relate to two different purposes: for students, on AI for completing coursework; and for instructors, on AI for creating coursework.

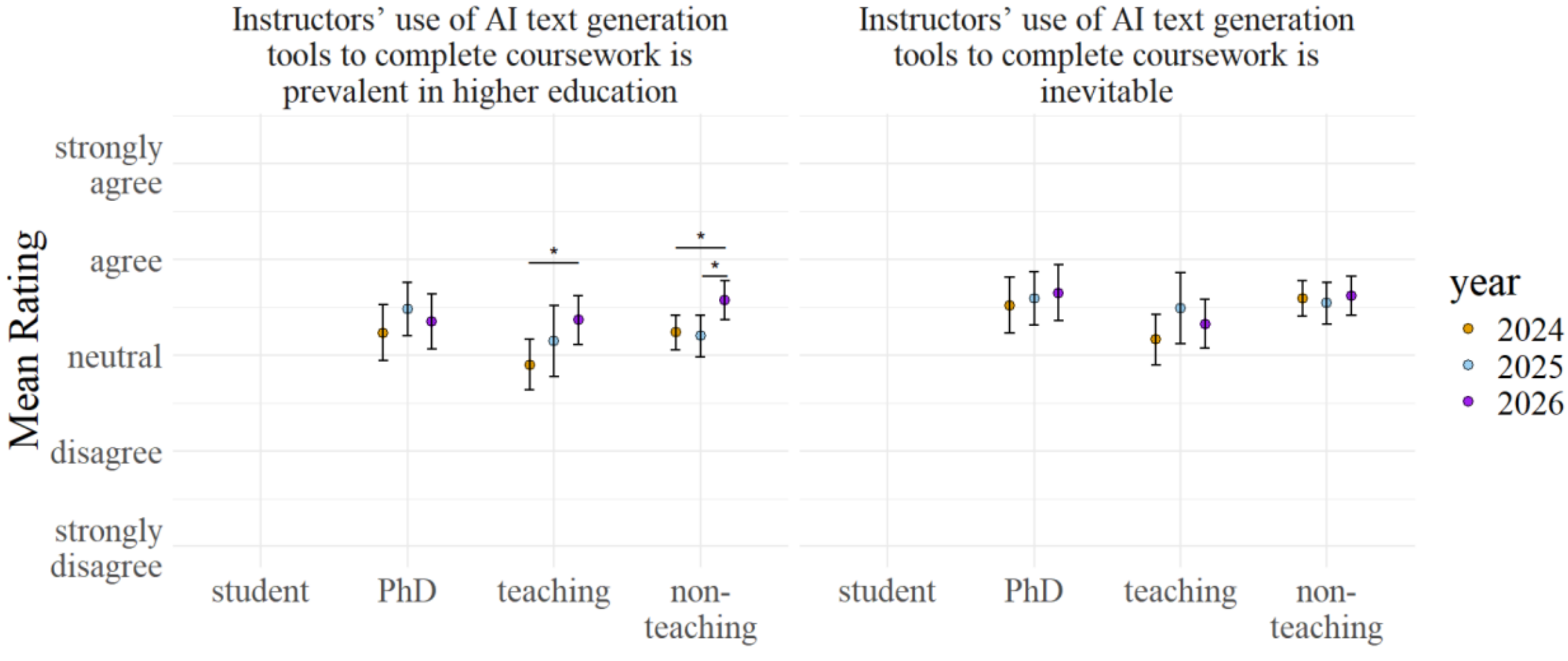

Figure 12

Figure 13

### 4.5 Institutional policy

Results for Institutional policy are presented in Figures 14-16. The ratings for the use of AI in marking are particularly striking: all four populations gave consistently low ratings across all time points. That is, essentially no one trusts AI for marking, and this trust has not changed (Figure 14). Across all timepoints, nearly half of respondents indicated “strongly disagree” (Appendix 1); of the remaining responses, most were “disagree”, with around 10% “agree” (2% of students also selected “strongly agree”). Moreover, all four populations generally agreed that the use of AI to complete coursework was inconsistent with academic integrity policies, with no significant changes observed across time points (Figure 16).

Nevertheless, the use of AI was not perceived as inappropriate in all contexts: perceptions were more positive across populations for the statement about the general benefits of using AI text-generation tools, across all timepoints (Figure 15).

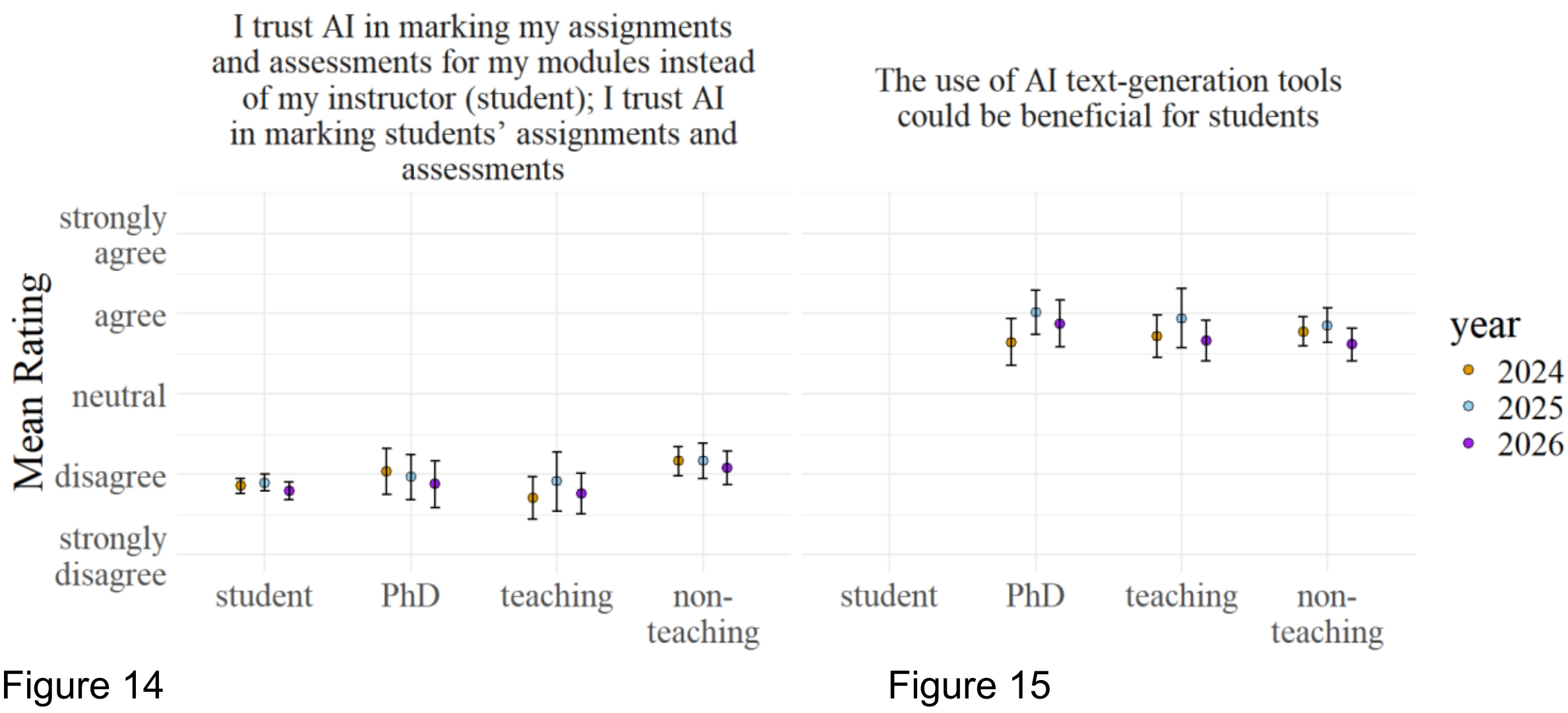

Figure 14

Figure 15

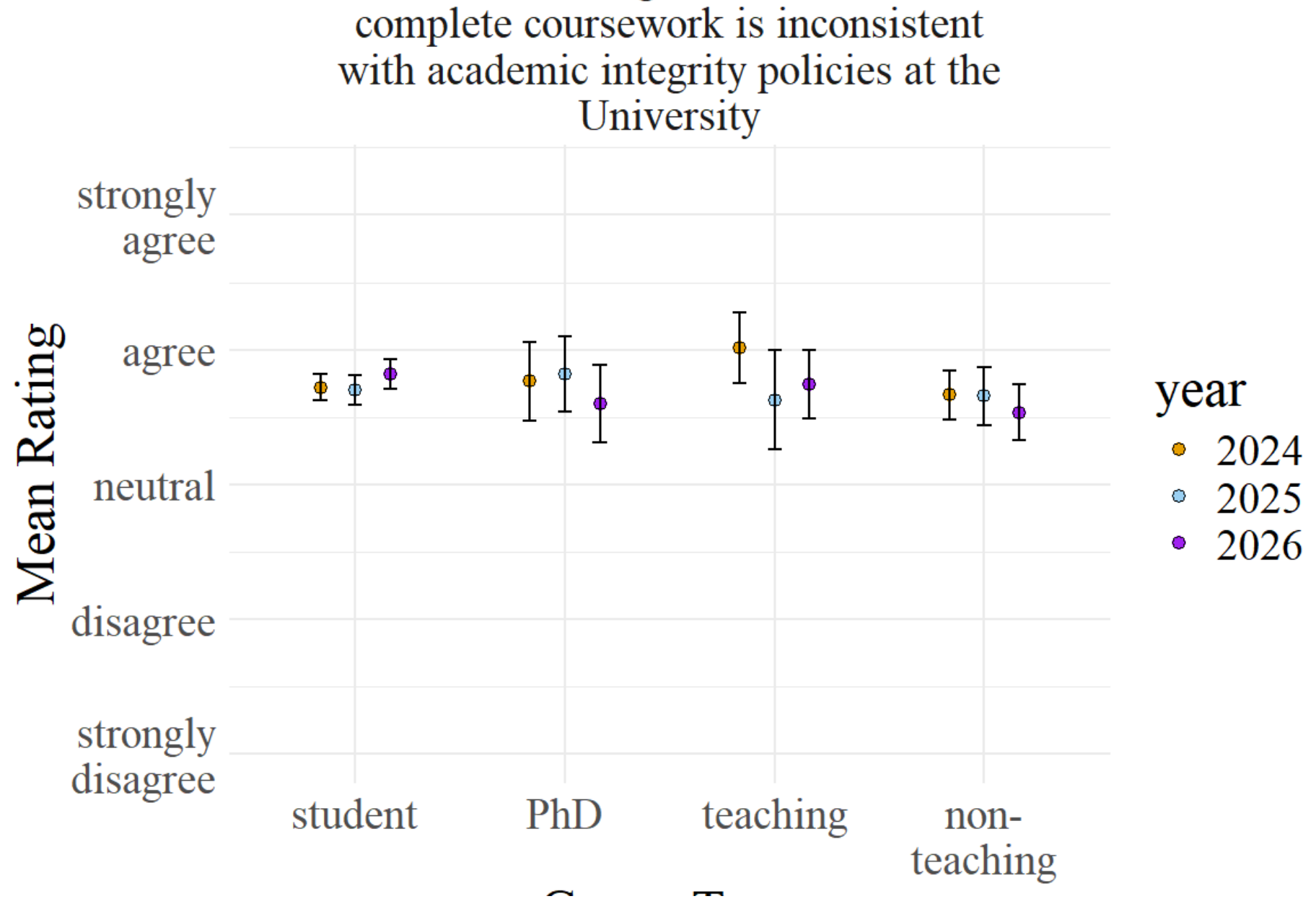

Figure 16

### 4.6 Student concerns (students only)

Finally, we present the ratings for statements presented to students only, relating to student concerns about AI. These included statements about the value and benefits of AI for students specifically, as well as statements about instructors' use of AI.

*4.6.1 Value/benefit*

The results for students' perceived value/benefit of AI in education contexts are presented in Figure 17. Together, they add some key perspective to the increases in experience and integration observed above: students' ratings decreased significantly for the value of AI in education, and for whether AI is used for "good and helpful reasons." Both trends are driven by a decreasing proportion of "agree" responses and an increasing proportion of "strongly disagree" responses (Appendix 1).

*4.6.2 Instructor use*

The results for students' perceptions of instructor use are presented in Figure 18. Collectively, they indicate limited confidence in instructors' AI use: students are not particularly confident in their instructors' use of AI, and their perception of AI-created resources has significantly decreased. Students also consistently indicated that they would want to be informed if their instructor was using AI resources on courses.

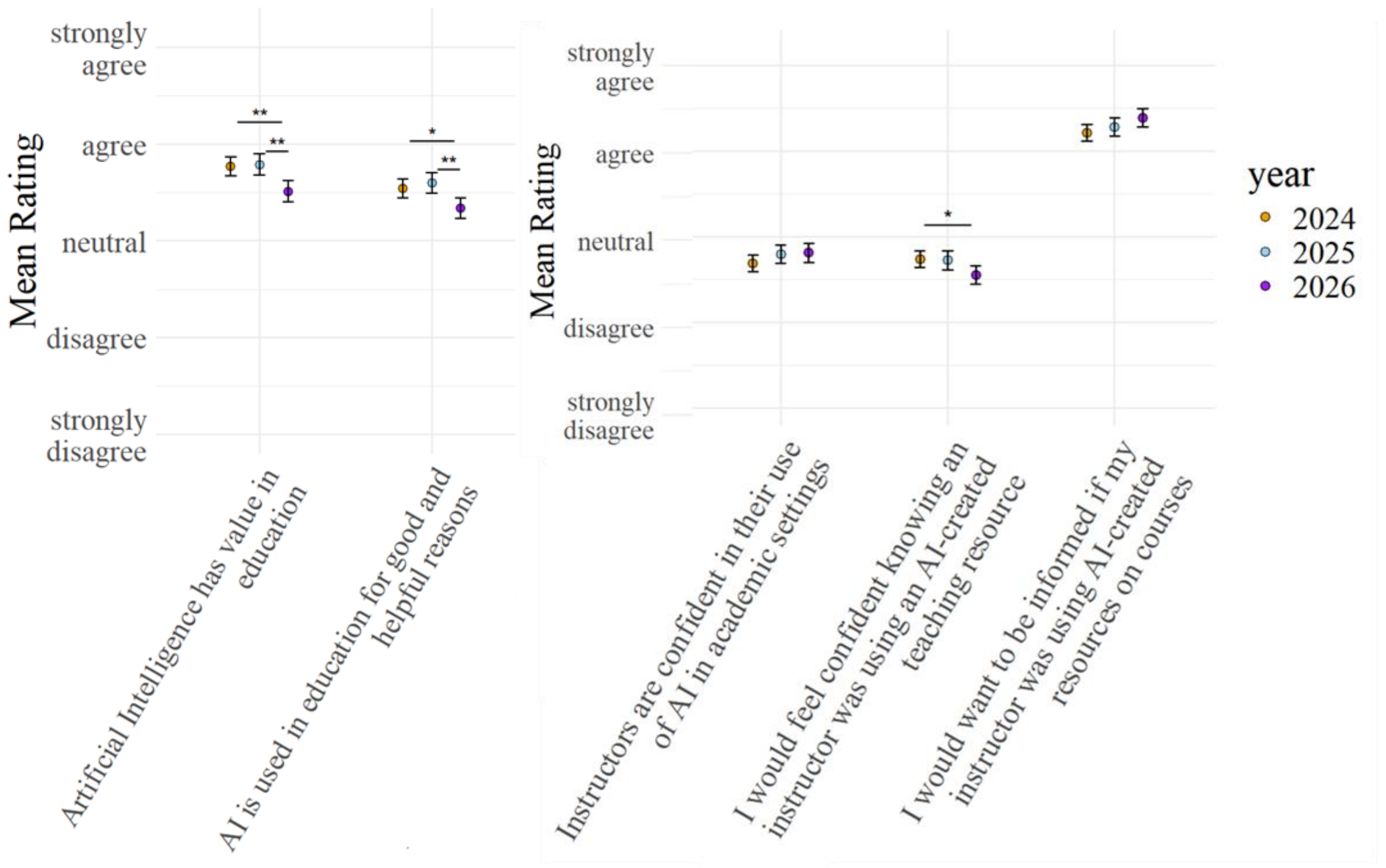

Figure 17 Figure 18

## 5. Discussion

This study investigated perceptions of AI - particularly Generative AI - in HE contexts, across different populations and over three different timepoints: 2024, 2025 and 2026. The results reveal a picture in which staff and students respond to AI with greater complexity than either uncritical acceptance or rejection would suggest. Generative AI has become widespread in academic practice. A general openness to further training persists across populations. Respondents nevertheless recognise that such use bears on academic integrity. Few regard these tools as suitable for complex tasks such as assessment or the design of teaching resources. Positive perceptions coexist with firm views about appropriate and inappropriate application. Marking offers the clearest instance of a function few respondents would entrust to AI. Caution about future use accompanies these views. One might venture that an absence of clear worked examples explains some of the hesitancy. Such a stance indicates critical reflection on the use of AI rather than reflexive endorsement or rejection. Awareness of this kind accords poorly with blanket generalisations, such as characterising AI as 'good and helpful'. As familiarity with these technologies grows, abstract discussion of AI ought to give way to consideration of specific applications on a case-by-case basis.

In this discussion section, we first consider potential sources of the observed results, and then relate the findings to a broader theoretical framework for interpreting staff and student perceptions (Gerard et al 2026). Finally, we discuss the implications for institutional policy, particularly in relation to evidence-based approaches.

**5.1 Summary of results**

The findings demonstrate clear changes in perceptions of generative AI across the three survey waves. In addition, differences between the four populations proved as informative as the changes over time. In the following sections, we revisit the results from each broader theme and consider potential sources and implications.

*5.1.1 Familiarity*

Familiarity and prior experience with generative AI rose significantly across the three survey waves for almost every group. Students, teaching staff, and non-teaching staff all reported greater familiarity with ChatGPT and similar tools by 2026 than in 2024, alongside increased experience of using them. The exception was familiarity among doctoral researchers, where the smaller sample size likely limited statistical detection of an effect already evident in the descriptive trends. Mean ratings sat consistently between ‘Agree’ and ‘Strongly Agree’ from the outset, yet the underlying distributions tell a more revealing story. Increases were driven by participants moving out of ‘Disagree’ and ‘Strongly Disagree’ categories rather than from neutral positions.

The shift suggests that those once unfamiliar with these tools became actively familiar rather than ambivalent. The pattern reflects a wider cultural shift in which generative AI has moved from a specialist novelty to a routine feature of academic life, normalised across student and staff populations alike. The implication for universities is that AI literacy can no longer be framed as an introductory orientation. Familiarity is now baseline, and institutional efforts may be better directed toward critical engagement, ethical use and discipline-specific applications rather than awareness-raising.

*5.1.2 Use in practice*

A striking contrast emerged between the actual integration of AI and future plans for its use. Students, teaching staff, and non-teaching staff all reported significant increases in addressing AI in their teaching or learning, with non-teaching staff moving from ‘Disagree’ in 2024 to a neutral position by 2026. Integration of AI tools in research, instruction or work also rose significantly across all groups, though staff ratings remained lower than student ratings throughout. Future plans to use AI, by contrast, showed only modest movement, with students nudging slightly upward and PhD, teaching and non-teaching staff remaining clustered around neutral.

The gap between current practice and stated future intent suggests that adoption is happening organically rather than through deliberate planning. AI use is becoming embedded in academic routines without explicit commitment. The finding raises questions about how universities can support sustainable, intentional integration rather than relying on incremental drift. If staff are already using these tools without planning to do so, institutional guidance needs to address actual practice as it stands, not the hypothetical adoption pathway envisaged in earlier policy documents.

*5.1.3 Training*

Reported participation in AI training rose significantly across all populations between 2024 and 2026, with students and teaching staff showing the largest gains. Interest in receiving further training, however, decreased slightly among students over the same period, with negative $\beta$ values recorded between 2024 and 2026. Staff interest in training remained steady across the period.

The rise in reported training reflects an institutional response to growing demand, as universities began offering workshops, modules and guidance documents in greater numbers from 2024 onward. The decline in student interest likely reflects a saturation effect: as students gained confidence through informal experimentation, the perceived need for formal instruction diminished. The continuing staff appetite for professional development suggests an opposite pattern, with staff still seeking structured support. The divergence between student and staff demand has implications for resource allocation. Universities may need to shift from generic AI literacy provision toward more advanced or discipline-specific offerings if training is to remain relevant for student users, while continuing to expand foundational training for staff.

*5.1.4 AI and coursework*

Perceptions of AI as a prevalent feature of higher education coursework strengthened significantly between 2024 and 2026 across students, teaching staff and non-teaching staff. Teaching staff showed the most pronounced shift, moving close to ‘Strongly Agree’ by 2026, while ratings of inevitability remained stable across all groups. Views on whether students should be restricted from using AI moved in opposite directions: students became less likely to support restriction over time, while teaching staff became more likely to do so between 2024 and 2025.

The divergence captures the widening gap at the heart of this study. Students treat the use of AI as a legitimate everyday practice, whereas teaching staff increasingly see it as something that requires boundaries. Acceptance of prevalence does not amount to acceptance of permissibility. Staff recognise that AI use is widespread without endorsing it, which has implications for how assessment design and academic integrity frameworks evolve in the coming years. The challenge for universities is to develop assessment practices that account for routine AI use without abandoning the expectations of independent thought and analysis that underpin higher education credentials.

#### *5.1.5 Institutional policy*

Trust in AI for marking assignments and assessments remained low across all populations and all three years, with mean ratings clustered around 'Disagree' for students, teaching staff and non-teaching staff. Doctoral researchers showed slightly higher ratings, though still below neutral. The belief that AI could be beneficial for students sat consistently around 'Agree' across all groups, with little year-on-year movement. Views on whether AI use is inconsistent with academic integrity policies showed more variation, with teaching staff moving from 'Disagree' in 2024 toward neutral by 2026.

The shift among teaching staff suggests growing recognition that institutional frameworks have struggled to keep pace with student practice. The persistence of low trust in AI marking, alongside acceptance of AI's potential benefit, points to a conditional acceptance: AI is welcomed as a learning aid but not as an evaluative authority. Institutional policy will need to address this distinction directly by separating questions of student use from those of staff or system use. Generic policy statements that treat AI as a single phenomenon are unlikely to capture the differentiated trust the data reveal.

#### *5.1.6 Student concerns*

Student perceptions of AI's value in education declined significantly between 2025 and 2026, alongside a parallel decline in agreement that AI is used in education for good and helpful reasons. Confidence in instructors' use of AI-created teaching resources also fell significantly over the same period. The pattern is unexpected, given the simultaneous rise in students' familiarity, experience, and integration with AI documented in earlier sections.

Increased exposure has not produced increased enthusiasm. Students appear to be developing a more critical stance as they encounter the limitations of these tools, the inconsistencies in how staff use them, or the broader cultural debate about AI's role in higher education. The trajectory complicates any simple narrative of normalisation. Familiarity does not equal endorsement, and routine use coexists with growing scepticism. Universities responding to the rise of AI need to address student concerns about quality and purpose, as well as questions of access and skills. The finding warrants further qualitative exploration to identify the specific experiences driving this shift.

### 5.2 Comparison with previous studies

These results align with earlier single-time-point studies that described either enthusiasm (Chan and Hu 2023) or concern (Alshamy et al. 2025) among different groups. However, unlike those snapshot studies, our longitudinal design captures how perceptions evolve. Initial novelty and uncertainty in 2024 gave way to routinised use by 2025, paralleling sector-wide surveys that reported rapidly increasing student adoption (Tyton Partners 2023; Contractor and Reyes 2025). Staff caution persisted, but our data show that this caution did not prevent students from incorporating AI into their everyday academic practice. This divergence is only partially visible in cross-sectional research; by tracking responses over time, our study makes the widening gap explicit. To place these changes in further context, we consider how they contribute to a broader framework for AI use in HE, focusing on the framework developed by Gerard et al. (2026) for AI in HE across populations.

### *5.2.1 Framework for AI perceptions in HE*

To develop a framework on AI in HE contexts, Gerard et al. (2026) conducted separate focus groups with teaching staff, non-teaching staff, PhD students, and students. Under this qualitative framework, three themes (T), each with sub-themes, were identified. T1: Attitudes to AI; prior experience, positive attitudes and negative attitudes. T2: Highlights & Challenges of AI Application in HE; opportunities, obstacles and optimising AI. T3: Future of AI; staff and student training.

#### 5.2.1.1 Theme 1: Attitudes to AI

Attitudes towards AI appeared to be shaped by prior experience, as those with greater experience reported more positive views and described a range of academic uses, e.g., an efficient search tool, a personalised learning resource, and a mentoring aid. Some participants limited AI use to idea generation and raised concerns about students' reliance on AI. All participants acknowledged the risk of inaccurate or spurious outputs, emphasising the importance of critical appraisal and fact-checking.

This growth of familiarity and experience extends the qualitative framework, as all groups previously discussed experiences with AI, especially the non-teaching staff and PhD and student groups. The current results indicate that this engagement has continued and grown over time. Interestingly, the teaching staff reported higher levels of experience in 2026 than students and non-teaching staff. The previous study showed that although the teaching staff had experience with AI, they held fewer positive attitudes than the other groups. This higher level of engagement among teaching staff may also indicate a shift in attitudes towards AI. Although attitudes were not directly measured in the current study, previous findings suggested that more positive attitudes towards AI were associated with greater levels of experience and use. It may be that enhanced exposure to AI has led to more acceptance among teaching staff.

Expanding on participant experiences, 'Use in Practice' showed that students felt that AI was being increasingly integrated into teaching practices across the 3 years (Figures 4-5). However, it is unclear whether these responses reflect the actual embedding of AI-related methods within formal teaching practices or the enhancement of self-directed learning. The previous qualitative findings suggest that students were already engaging with AI in their studies more than the teaching staff were in their teaching. The observed increase might therefore reflect growing confidence and familiarity with AI among students, educators, or both, leading to greater engagement and discussion of AI in teaching activities.

Despite this, participant groups provided 'neutral' responses across the 3 time points for planning to use, integrate AI in coursework, research, instruction or jobs (Figure 6). Perhaps, this may be partly explained by participants' awareness of the limitations and risks associated with AI. The previous study highlighted that all participants had a clear understanding of AI-generated inaccuracies, leading to concerns about overreliance and the need to check facts. For example, negative attitudes were evident when participants discussed observing AI-generated content in others' work, suggesting that these users failed to critically review or edit the content. It may be possible that such concerns do not translate into significant intentions to integrate AI into coursework, teaching and research. Therefore, these neutral responses might reflect informed, cautious attitudes in which participants acknowledge the benefits of AI whilst remaining guarded about the risks. This may raise awareness of the ethical and practical challenges brought about by AI.

5.2.1.2 Theme 2: Highlights & Challenges of AI Application in HE

Participants discussed the dual potential of AI to support and hinder university education, highlighting its value in improving accessibility and support while stressing the need for ethical use and transparency. Concerns were raised by all groups regarding the use of AI in assessment, with agreement that a deep understanding of AI's limitations was needed to optimise its future use.

Increased perceptions of AI as a prevalent feature of coursework and agreement on the inevitability of AI use align with the qualitative framework, in which most participants viewed AI as providing opportunities rather than obstacles. Participants shared their experiences of benefits, including tailored learning pathways, increased accessibility, and academic mentoring. These insights may help explain why, in the current analysis, student use of AI in coursework is thought to be increasingly common. It may be that students recognise the use of AI as acceptable, thereby increasing their comfort level in integrating AI into their academic work. Perhaps it is seen as a valuable learning tool and a routine method for completing coursework. However, the findings may also indicate important differences between students and teaching staff. While the teaching staff acknowledged some benefits, including discussions under this theme and T1, they also discussed potential challenges, including a lack of academic progression and impacts on 'deeper learning'. This may also reflect greater awareness of AI use in student coursework, including exposure to AI-generated content, hallucinations, or inaccurate information. It may be that participants recognise the benefits of AI when it is used appropriately, yet are aware of its potential to hinder student learning. Interestingly, students' views on restricting AI in coursework lessened over time, while teaching staff felt that they could increase. Overall, this may reflect the normalisation of AI within higher education for all staff and students, alongside the perceived awareness of the challenges associated with its use from a teaching perspective.

Teaching and non-teaching staff reported increased perceptions of instructors' AI use over time, suggesting increased engagement with AI in assessment. By comparison, PhD researchers reported no significant changes, which may indicate that staff have greater visibility into their colleagues' assessment practices than researchers do. Within the qualitative framework (T2 and T3), all participants agreed that AI is 'here to stay' and, as such, the current perceived increase may also suggest that teaching staff are exploring AI's potential for coursework development and for reducing their administrative workload. Despite these increases in perceived use, scores regarding the inevitability of instructors' use of AI remained stable. Within the qualitative framework, instructors' use did not extend beyond a potential tool for teaching and coursework development, and it appears to have been maintained as a supportive option rather than a replacement for their professional capabilities. It may be that students utilise AI models for help with coursework and for generating written content, making their use more evident. Comparatively, instructors may use them to draft learning and assessment materials and to support administrative tasks, which are not always evident to students and other staff.

Within the qualitative framework, participants emphasised that AI's optimisation was contingent on managing its opportunities whilst mitigating known risks. For example, AI could be used as a scaffolding tool to help with brainstorming and structuring work, while being cognisant of its limitations and triangulating with reliable sources. The current decline in students' perceptions of AI's value and benefits aligns with these earlier thoughts. Although students have become more familiar with AI and adopted it in their academic work, they may now be more aware of its limitations in terms of accuracy and reliability. This increased exposure may have produced more criticism, rather than passive acceptance. Equally, reduced confidence in instructors' use of AI-generated teaching resources may reflect students' growing awareness of these limitations, and they might question whether such materials meet the standards of quality, expertise, and transparency expected in higher education. The finding that students want to be informed when AI-generated resources are used supports the importance of transparency identified in the framework. Taken together, increased familiarity with AI does not necessarily lead to greater trust or perceived value. Rather, more experience may have deepened their critical understanding of AI's benefits and limitations. Therefore, the optimisation of AI in higher education should focus on promoting AI literacy and technical skills and, from an ethical standpoint, ensuring transparent and responsible use by staff and students.

#### 5.2.1.3 Theme 3: The Future of AI

Rather than imposing restrictions, participants emphasised the need for clear institutional policies, guidelines and detection tools to support responsible AI use in higher education. Transparency was viewed as fundamental for maintaining trust, academic integrity, data protection and intellectual property rights. Effective AI training for staff and students was considered vital, with suggestions that it should be embedded within academic programmes and informed by industry stakeholders.

The qualitative framework provides some context for understanding the current changes in demand and future interest in AI training. Under the framework, all participants called for training that incorporated real-world, practical applications developed and supported at the institutional level. The current analysis appears to indicate that these recommendations were implemented, as participation in AI training increased across all groups, particularly among teaching staff and students. However, the demand is not straightforward, as student interest in further training has decreased despite increased participation. Perhaps this suggests that introductory training, coupled with their prior and ongoing experience, was satisfactory for some students but not for all. It may indicate that the available training was below their level of prior experience and did not enhance their developing skills, making the training irrelevant or repetitive.

Comparatively, staff interest in training remained stable despite increased participation, which may reflect findings from the qualitative framework, as staff wanted training that was short and easy to digest due to time and workload constraints. It may be that staff did not have the time to engage with all available training. Alternatively, this continued demand for training may reflect the rapid growth of AI technologies, coupled with the lack of AI detection models, to be managed alongside professional concerns, such as inappropriate use in education. Students may have enhanced their AI skills and knowledge through regular use, whereas staff may need the safety of ongoing professional development to ensure AI is used effectively while maintaining quality assurance and academic integrity. Overall, this may highlight a need for more distinct training pathways that reflect varying levels of experience, and which is tailored to the user's specific needs. Such an approach may help ensure that training remains relevant as AI becomes increasingly embedded within higher education practice.

### 5.3 Implications for future policy and training

The changing nature of perceptions has significant implications for universities. First, institutional policy cannot be static: guidance written for 2023 is already outdated in 2025. Universities need to develop adaptive frameworks that respond to new uses and emerging risks. Crucially, such frameworks must take an evidence-based approach to AI integration. For example, while all populations were generally open to receiving further training, ratings were lower for plans to integrate AI into day-to-day activities. Moreover, these ratings remained stable across all time points, despite across-the-board increases in AI experience. Thus, the evidence suggests that greater experience with AI is not an indicator of increased plans to integrate AI; therefore, greater access to AI, even when accompanied by training, does not guarantee AI integration.

This raises the question of why plans to integrate AI did not increase and what influences them. Institutional policy must take these factors into account; thus, further research is needed to identify which factors affect plans to integrate AI and whether these vary across populations.

The evident gap between student practices and staff expectations underscores the urgency of this research to develop targeted training for teaching staff, not only on how AI might be used responsibly but also on understanding the forms of use already embedded in student learning. The inclusion of doctoral and non-teaching staff also highlights the importance of addressing the whole university community in policy development, ensuring that AI literacy and ethical practice extend beyond classrooms into research and professional services.

## 6. Conclusion

By documenting change over time, this study moves beyond descriptive accounts of perceptions of AI. It shows that attitudes are fluid, that institutional policy must evolve accordingly, and that staff training is essential if universities are to bridge the widening gap between official guidance and everyday academic practice.

**Data availability statement**

The raw data supporting the conclusions of this article will be made available by the authors, without undue reservation.

**Conflict of Interests**

The authors declare none

**Generative AI statement**

The authors declare that generative AI was not used in the creation of this manuscript.

**Appendix 1**

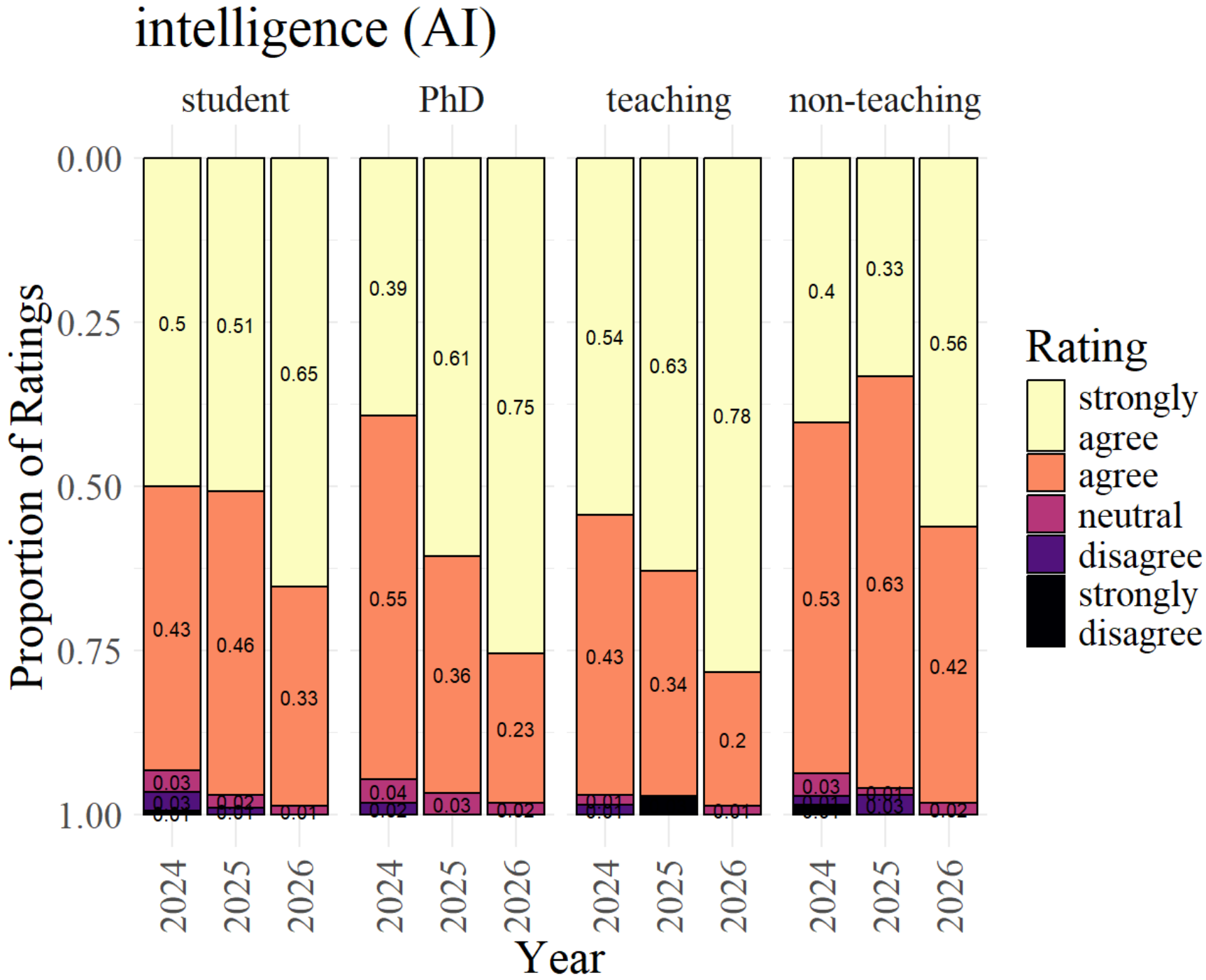

I am familiar with the concept of artificial intelligence (AI)
student
PhD
teaching
non-teaching
Proportion of Ratings
Year
Rating
strongly agree
agree
neutral
disagree
strongly disagree

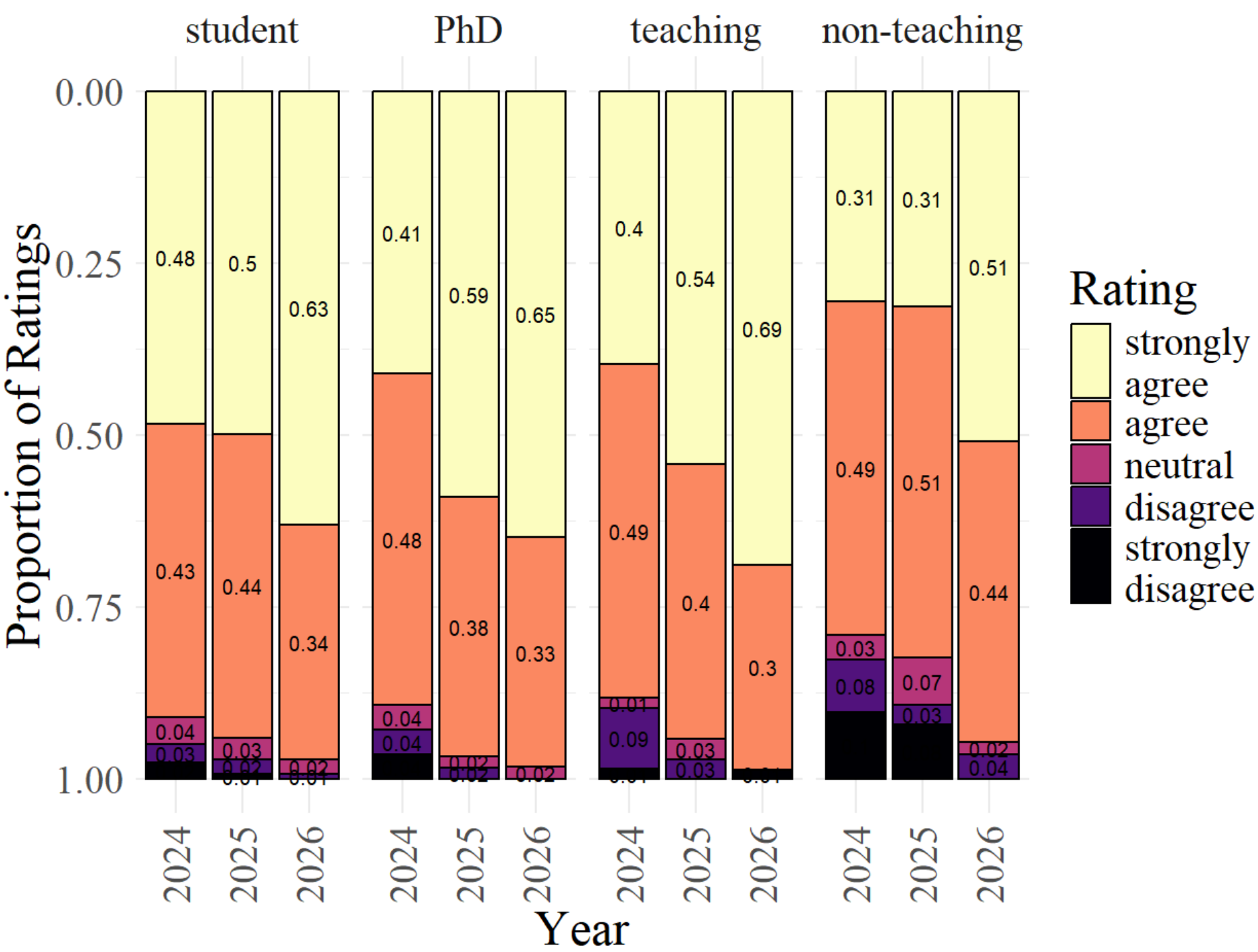

I am familiar with ChatGPT or other AI text generation tools
student
PhD
teaching
non-teaching
Proportion of Ratings
Year
Rating
strongly agree
agree
neutral
disagree
strongly disagree

I have experience using ChatGPT or other text generation tools

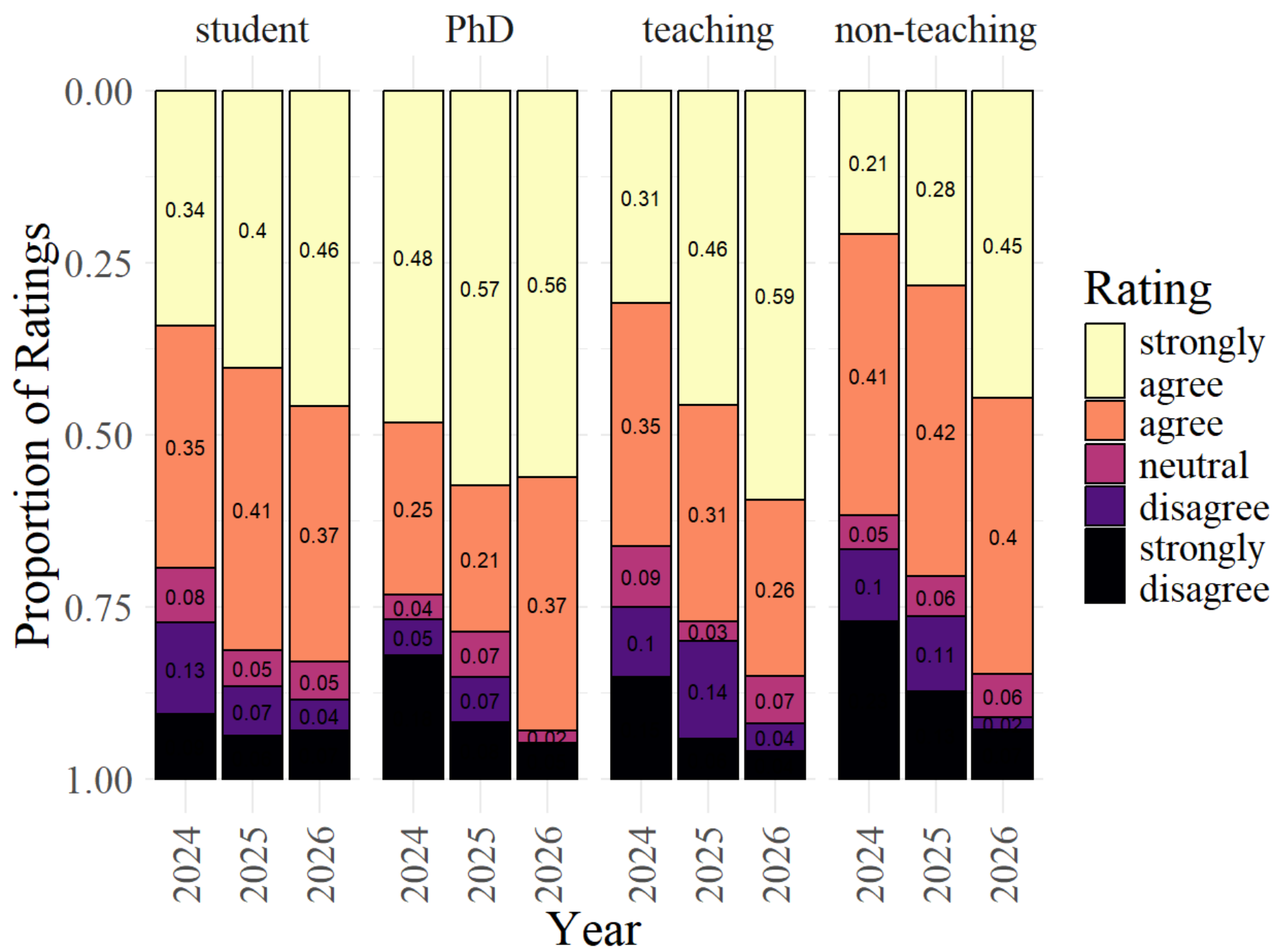

My instructors have addressed the use of AI in my modules (student); I have addressed the use of AI with my supervisor/students/team

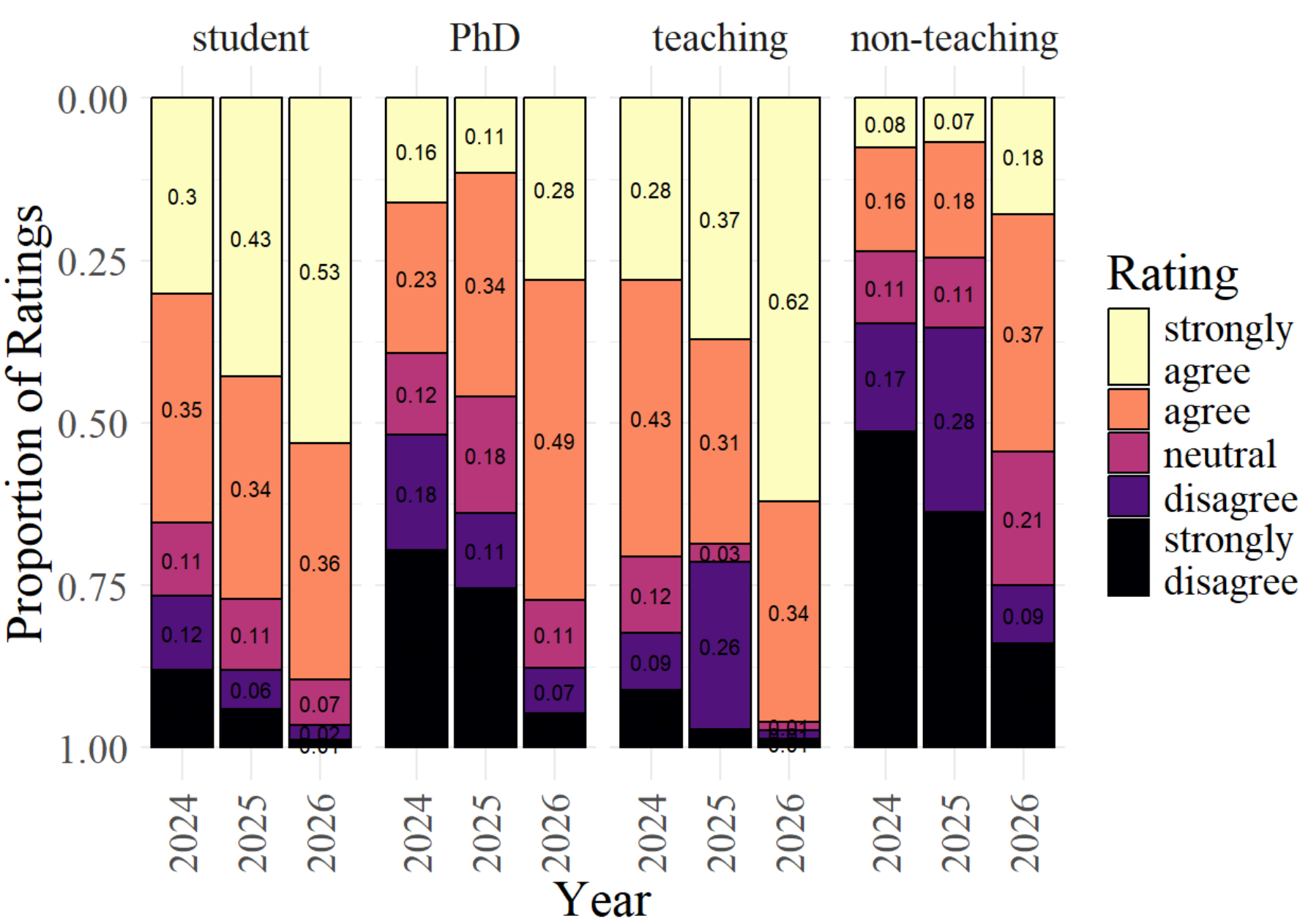

My instructors have integrated AI generators like ChatGPT into their instruction (student); I integrate AI tools in my research/instruction/job

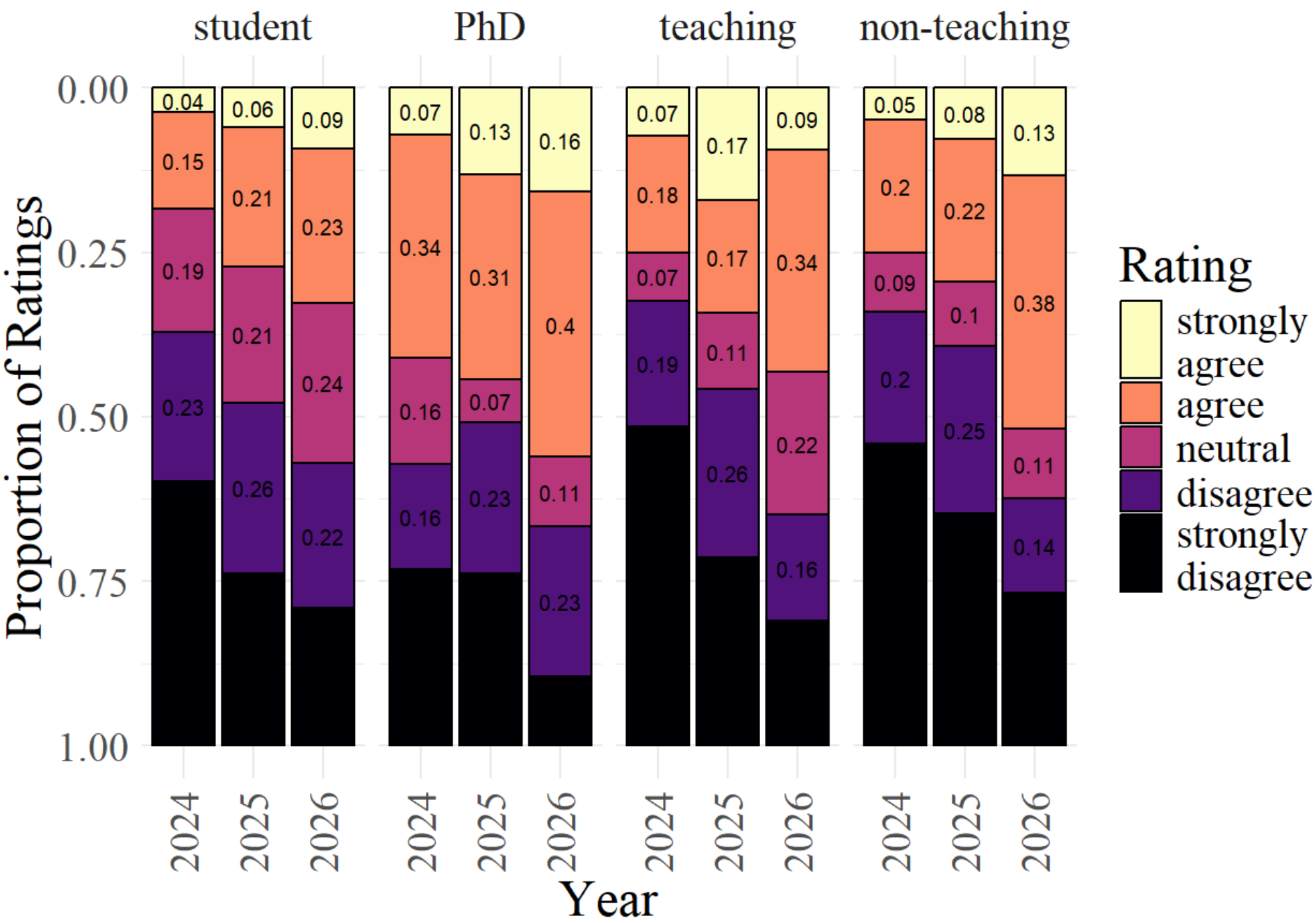

I plan to use ChatGPT or similar tools for my coursework in the future (student); I plan to integrate ChatGPT and/or other AI text or image generators into my research/instruction/job in the future

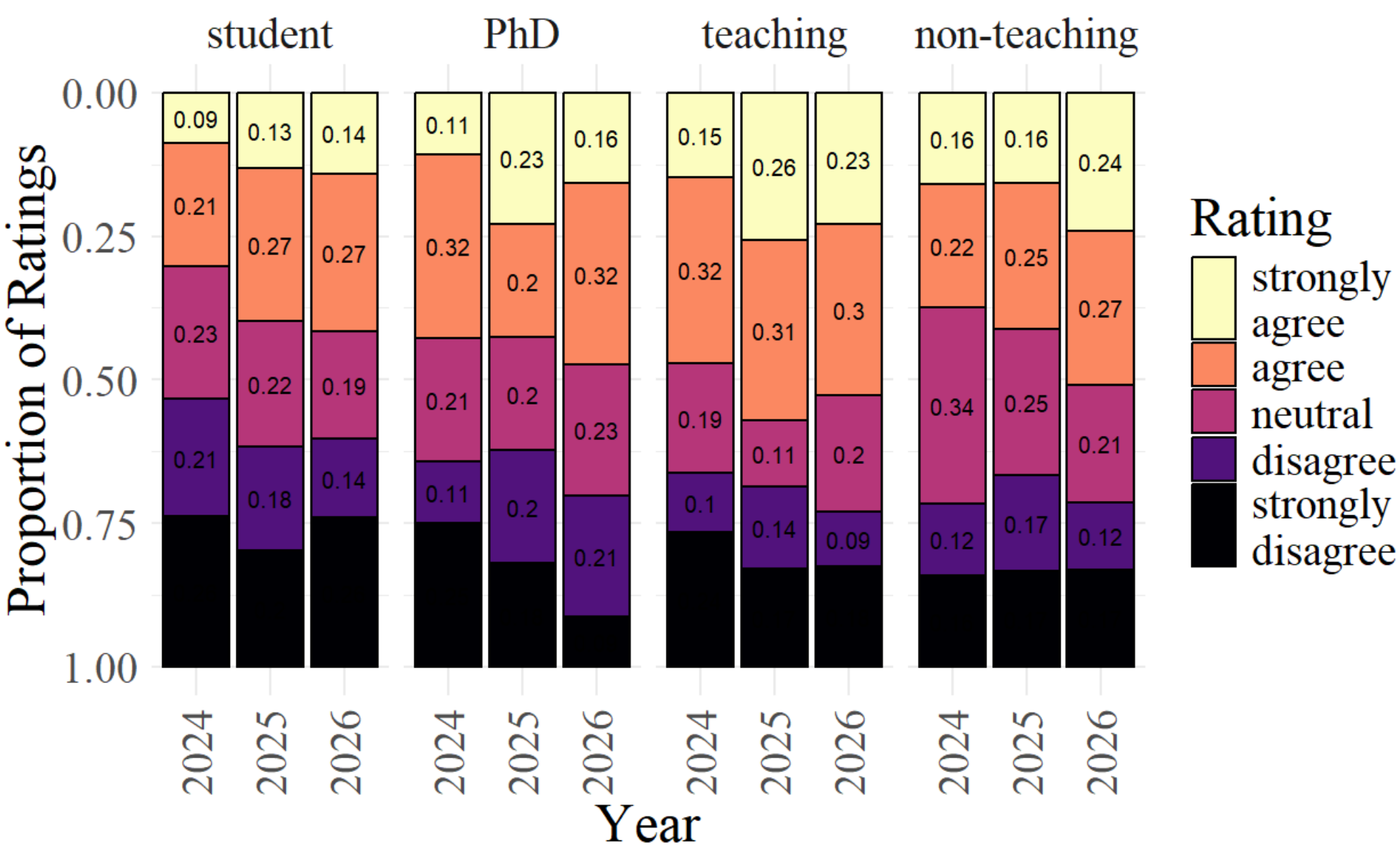

I have received instructions about how to use ChatGPT or similar tools (student); I have participated in training on the use of ChatGPT and/or similar AI text and image generators

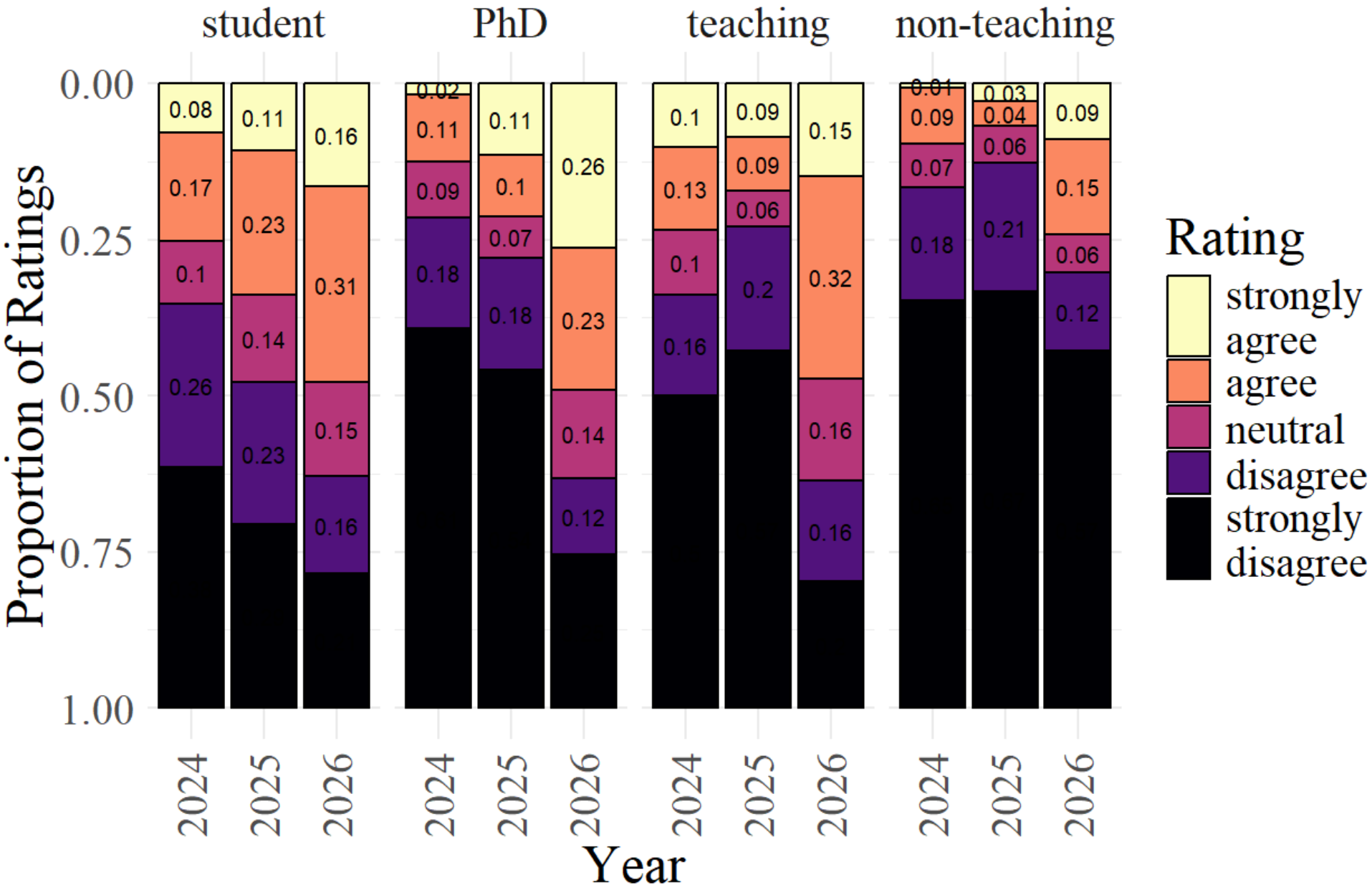

I would be open to receiving instructions about how to use ChatGPT or similar tools (student); I am interested in receiving training on the use of ChatGPT and/or similar AI text and image generators

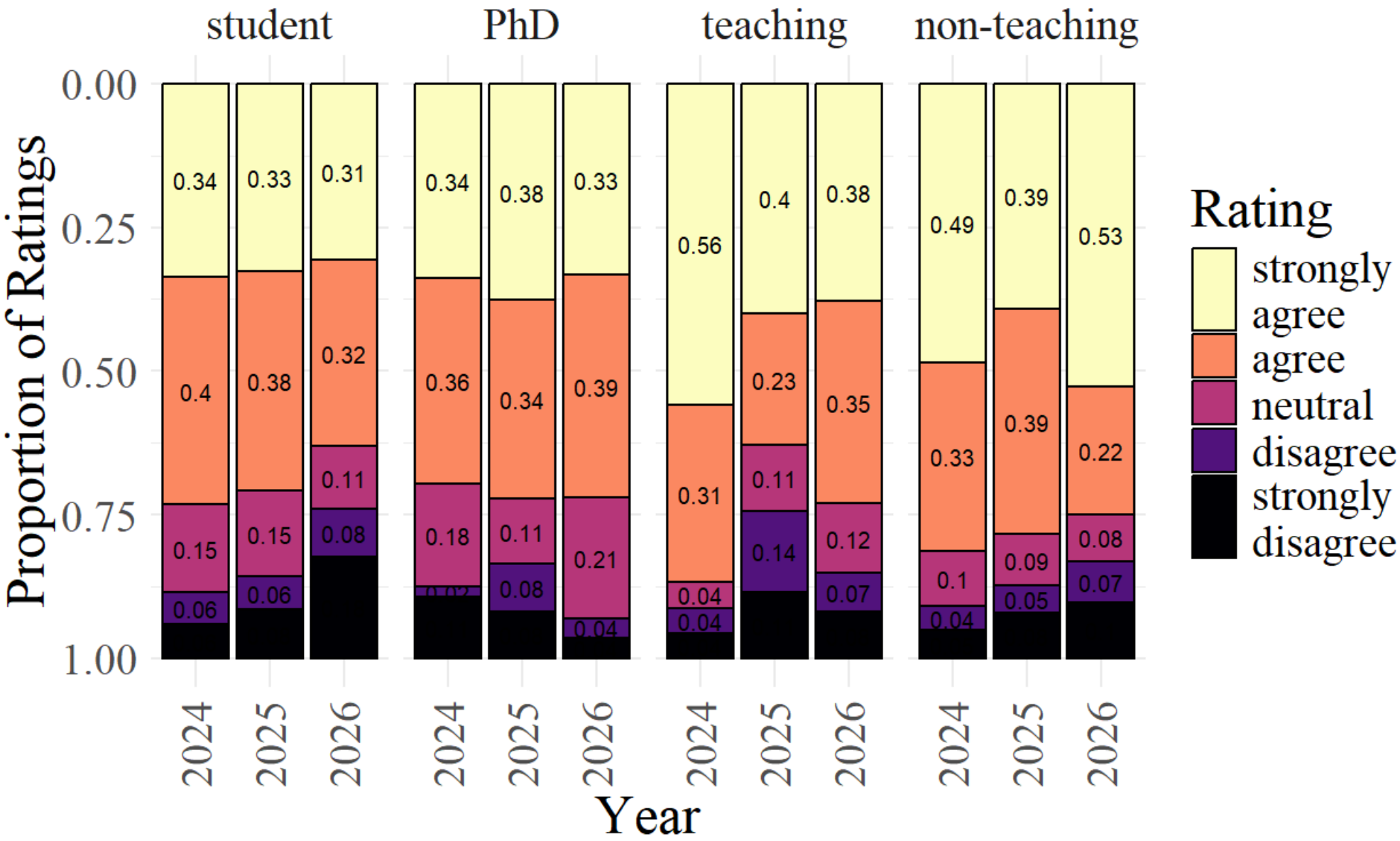

Students’ use of AI text generation tools to complete coursework is prevalent in higher education

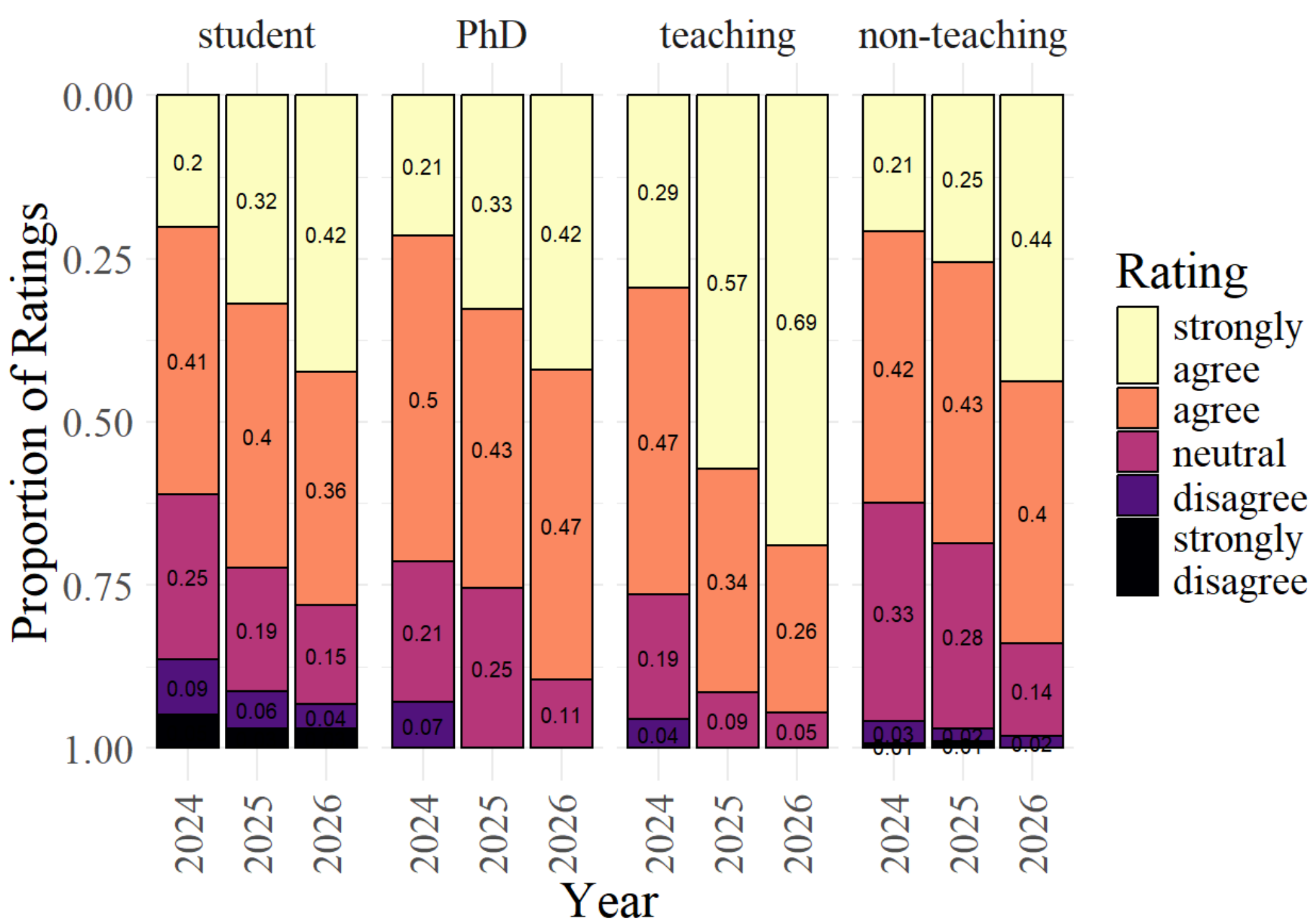

Students’ use of AI text generation tools to complete coursework is inevitable

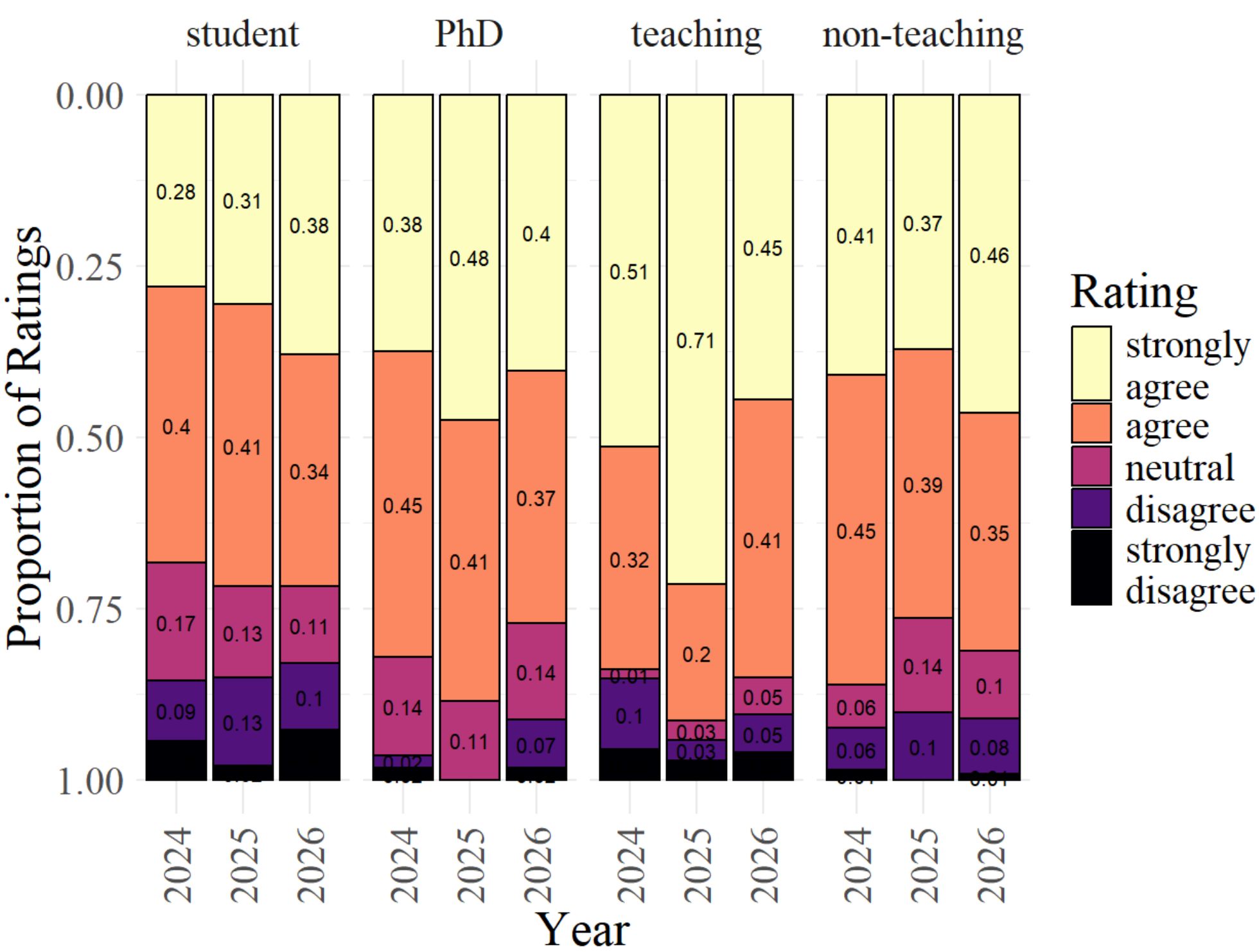

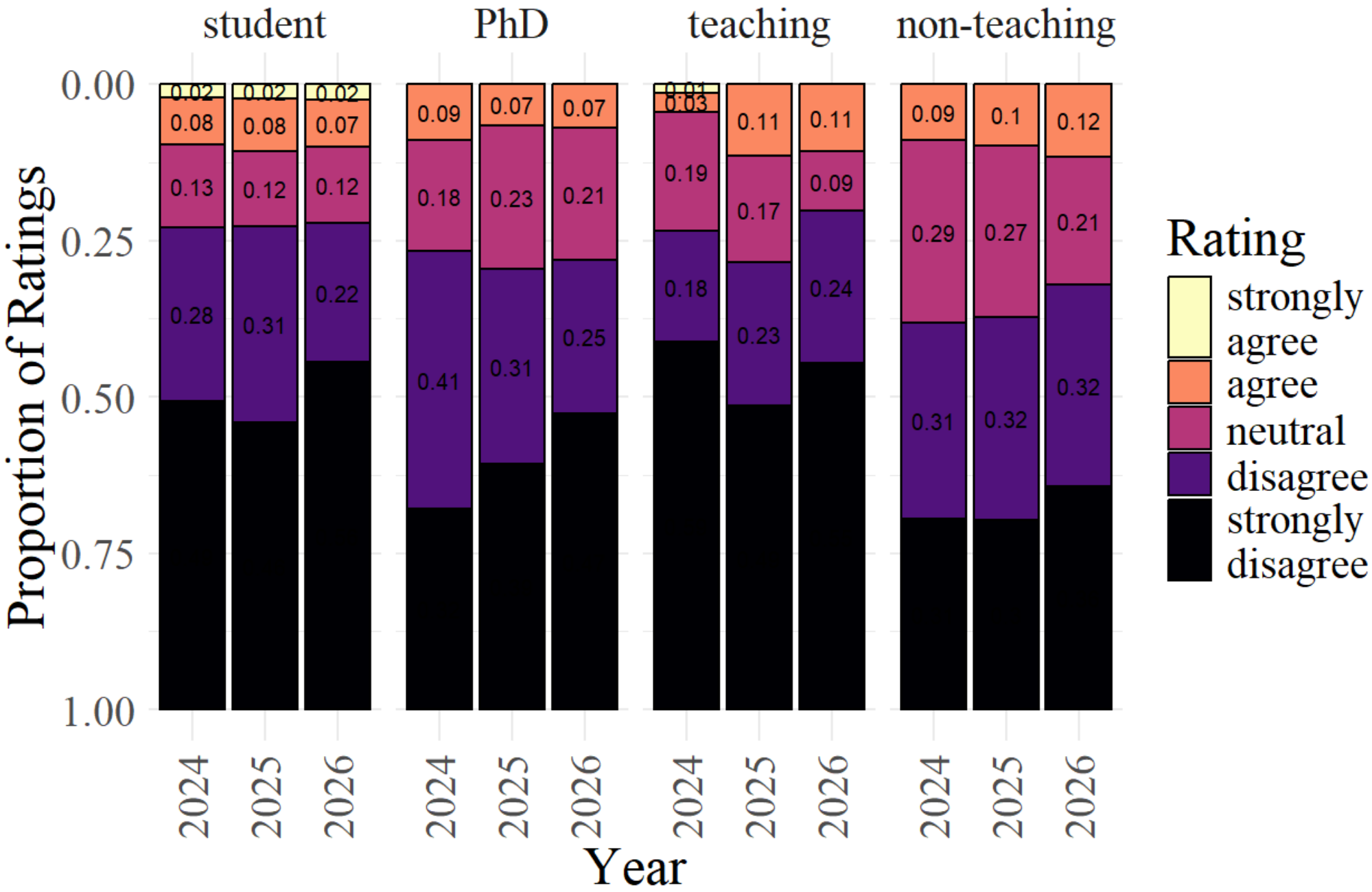
I trust AI in marking my assignments and assessments for my modules instead of my instructor (student); I trust AI in marking students' assignments and assessments
student
PhD
teaching
non-teaching
Proportion of Ratings
Year
Rating
strongly agree
agree
neutral
disagree
strongly disagree

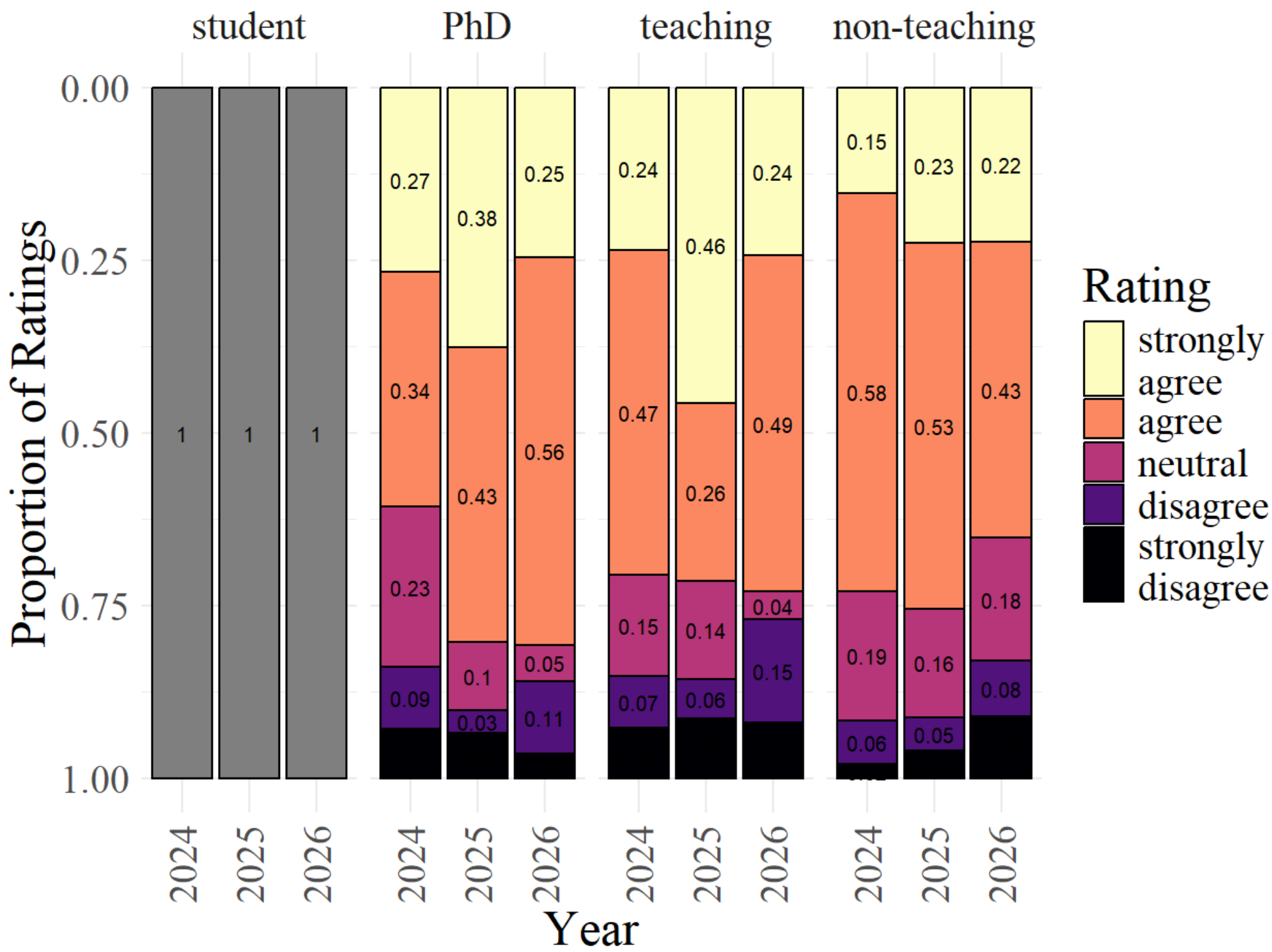
The use of AI text-generation tools could be beneficial for students
student
PhD
teaching
non-teaching
Proportion of Ratings
Year
Rating
strongly agree
agree
neutral
disagree
strongly disagree

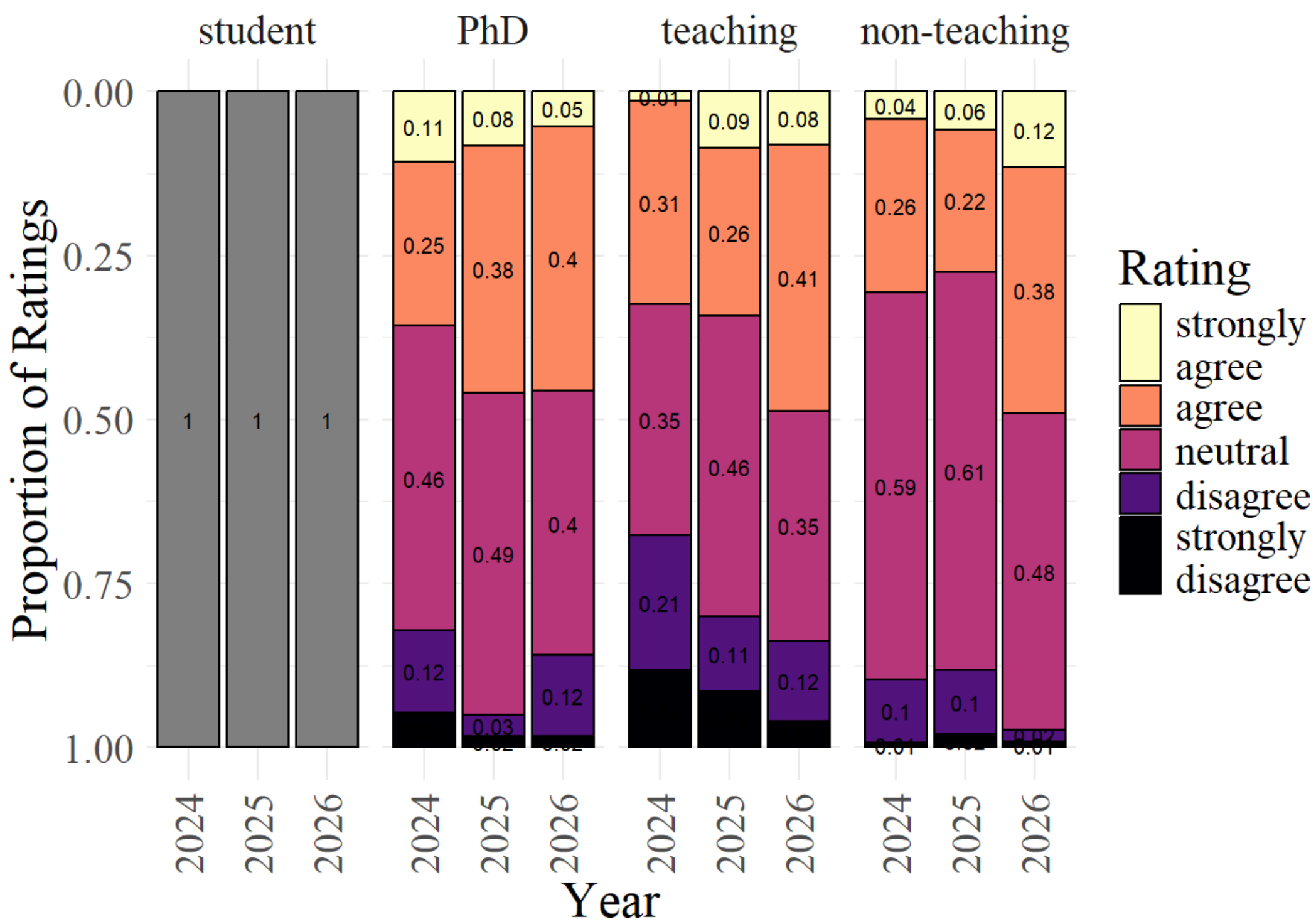
Instructors' use of AI text generation tools to complete coursework is prevalent in higher education
student
PhD
teaching
non-teaching
Proportion of Ratings
Year
Rating
strongly agree
agree
neutral
disagree
strongly disagree

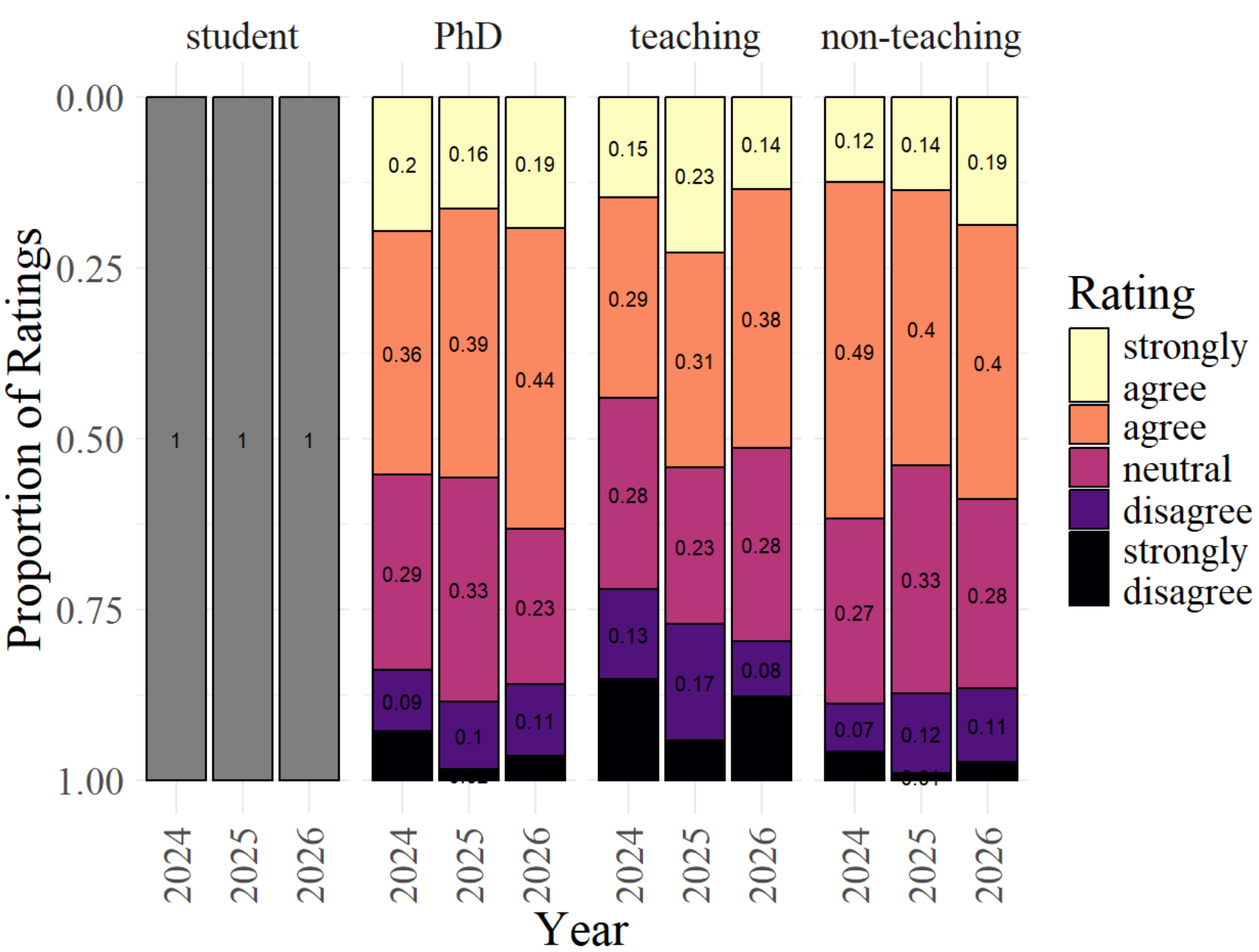
Instructors' use of AI text generation tools to complete coursework is inevitable
student
PhD
teaching
non-teaching
Proportion of Ratings
Year
Rating
strongly agree
agree
neutral
disagree
strongly disagree

Use of AI text generation tools to complete coursework is inconsistent with academic integrity policies at the University

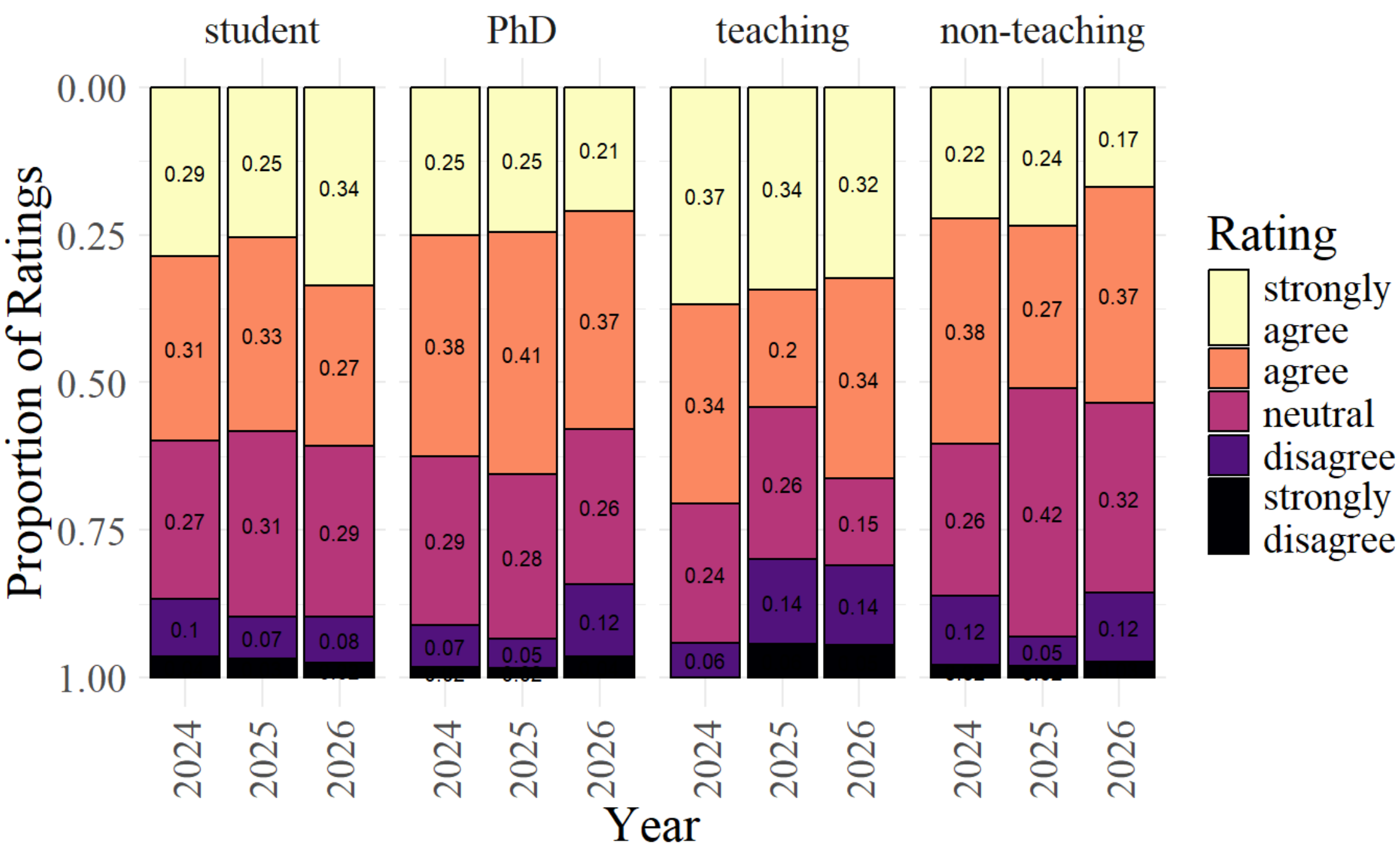

Students should be restricted from using AI for coursework

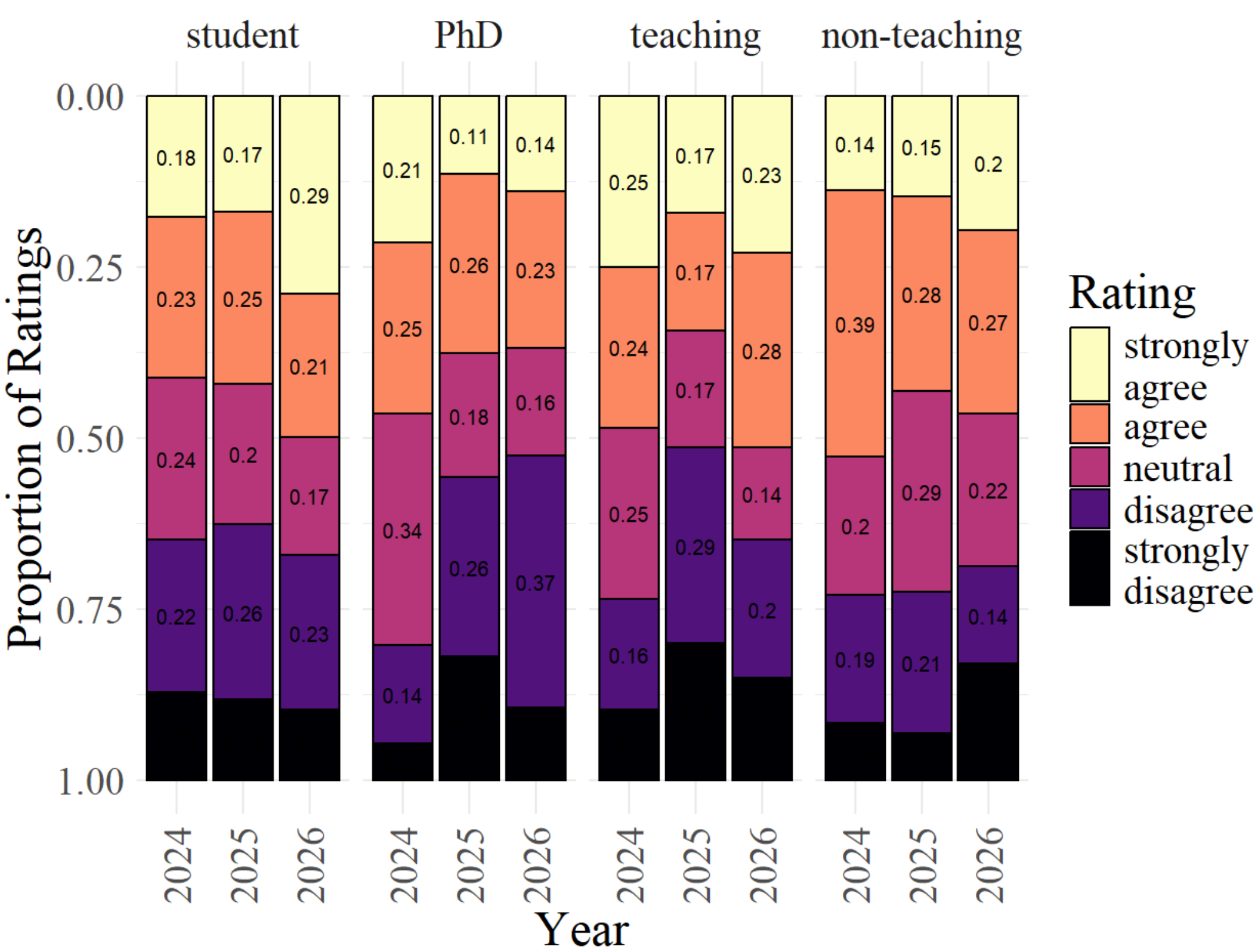

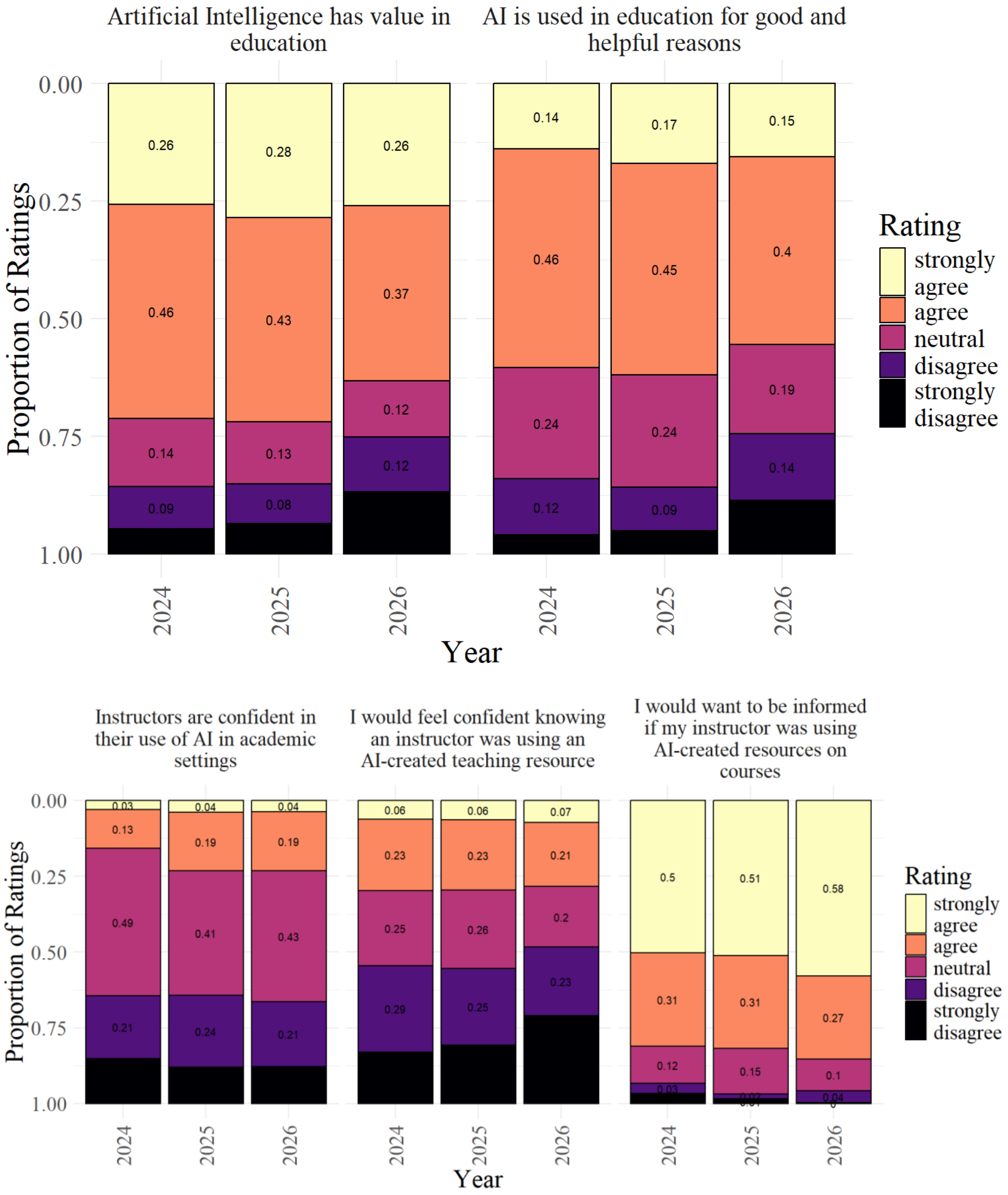
Artificial Intelligence has value in education
AI is used in education for good and helpful reasons
Proportion of Ratings
Year
Rating
strongly agree
agree
neutral
disagree
strongly disagree
Instructors are confident in their use of AI in academic settings
I would feel confident knowing an instructor was using an AI-created teaching resource
I would want to be informed if my instructor was using AI-created resources on courses